\documentclass[final,10pt,journal,letterpaper]{IEEEtran}

\IEEEoverridecommandlockouts

\usepackage{stmaryrd}
\usepackage{arydshln}
\usepackage{graphicx}
\usepackage{amsmath}
\usepackage{amssymb, verbatim}
\usepackage{mathrsfs}
\usepackage{dsfont}
\usepackage{url}
\usepackage[compress]{cite}
\usepackage{mdwlist}
\usepackage{paralist}
\usepackage{algorithmic}
\usepackage{algorithm}
\usepackage{tikz}
\usepackage{tkz-graph}
\usepackage[utf8]{inputenc}
\usepackage{multirow}
\usepackage{color}

\PassOptionsToPackage{usenames,dvipsnames,svgnames}{xcolor}
\usetikzlibrary{arrows,positioning,automata,calc}

\newtheorem{lemma}{\sc Lemma}[section]
\newtheorem{theorem}[lemma]{\sc Theorem}
\newtheorem{definition}[lemma]{\sc Definition}
\newtheorem{proposition}[lemma]{\sc Proposition}
\newtheorem{corollary}[lemma]{\sc Corollary}

\newtheorem{remark}{\sc Remark}[section]
\newtheorem{assumption}{\sc Assumption}[section]
\newtheorem{example}{\sc Example}

\renewcommand{\matrix}[2]{\left[\begin{array}{#1} #2 \end{array}\right] }

\DeclareMathOperator*{\argmin}{arg\,min}

\DeclareMathOperator*{\spans}{span}

\DeclareMathOperator*{\trace}{trace}

\renewcommand{\footnoterule}{%
  \kern -2pt
  \hrule width 0.3\textwidth height .5pt
  \kern 2pt
}

\definecolor{OliveGreen}{rgb}{0,0.5,0}

\newcommand{\N}{\llbracket N\rrbracket}

\newcommand{\V}{V}
\newcommand{\U}{U}

\renewcommand{\S}{\mathcal{S}}

\definecolor{lightgray}{rgb}{.80,.78,.69}

\newcommand{\farhad}[1]{{\color{black}#1}}

\begin{document}

\author{
Farhad~Farokhi,~Andr\'{e}~M.~H.~Teixeira,~and~C\'{e}dric~Langbort
\thanks{F. Farokhi is with the Department of Electrical and Electronic Engineering, University of Melbourne, Parkville, Victoria 3010, Australia. Email:farhad.farokhi@unimelb.edu.au} 
\thanks{A. Teixeira is with ACCESS Linnaeus Center, School of Electrical Engineering, KTH Royal Institute of Technology, Stockholm, Sweden. Email:andretei@ee.kth.se}
\thanks{The work of F.~Farokhi and A. Teixeira was supported by grants from the Swedish Research Council and the Knut and Alice Wallenberg Foundation. \farhad{F.~Farokhi was also supported by the Australian Research Council (LP130100605).}}
\thanks{C.~Langbort is with the Department of Aerospace Engineering, University of Illinois at Urbana–Champaign, IL, USA. Email:langbort@illinois.edu}
\thanks{The work of C.~Langbort was supported by grants from US Air Force Office of Scientific Research (AFOSR) under grant number MURI FA 9550-10-1-0573, and the US National Science Foundation under award \#1151076.}
\thanks{The authors would like to thank the hospitality of University of California at Berkeley, specifically, Alexandre M. Bayen and his research group, as well as CITRIS and CALTRAN, where they spent time when working on this paper. In addition, F.~Farokhi would like to thank Karl H. Johansson for discussions. \farhad{The paper was further greatly improved thanks to valuable comments from Serdar Y\"{u}ksel, Emrah Akyol, and Joel Sobel.} }}

\title{ Estimation with Strategic Sensors \thanks{A preliminary version of this paper was presented at the American Control Conference (ACC), 2014~\cite{FarokhiReport2013}.}}

\maketitle

\begin{abstract} We introduce a model of estimation in the presence of strategic, self-interested sensors. We employ a game-theoretic setup to model the interaction between the sensors and the receiver. The cost function of the receiver is equal to the estimation error variance while the cost function of the sensor contains an extra term which is determined by its private information. We start by the single sensor case in which the receiver has access to a noisy but honest side information in addition to the message transmitted by a strategic sensor. We study both static and dynamic estimation problems. For both these problems, we characterize a family of equilibria in which the sensor and the receiver employ simple strategies. Interestingly, for the dynamic estimation problem, we find an equilibrium for which the strategic sensor uses a memory-less policy. We generalize the static estimation setup to multiple sensors with synchronous communication structure (i.e., all the sensors transmit their messages simultaneously). We prove the maybe surprising fact that, for the constructed equilibrium in affine strategies, the estimation quality degrades as the number of sensors increases. However, if the sensors are herding (i.e., copying each other policies), the quality of the receiver's estimation improves as the number of sensors increases. Finally, we consider the asynchronous communication structure (i.e., the sensors transmit their messages sequentially).
\end{abstract} 

\section{Introduction} 
Over the past few years, a number of new technologies and concerns have made it necessary to consider problems related to estimation with self-interested and/or strategically deceitful sensors. 

One such technological concept is crowd- or participatory sensing where participants are relied upon (and sometimes actively recruited and incentivized) to sample and measure their environment, or provide personal information to be pooled and mined ``for the greater common good''. Examples include Sensorly for generating wireless network coverage maps~\cite{Sensorly} and Waze for traffic monitoring~\cite{Waze}.

In this case, strategic misreporting may occur for privacy reasons (if, e.g., a user is forced to report personal conditions to a participatory sensing system to get medical coverage, but does not trust the system enough to tell the truth), as a stealthy attack on the system (e.g., a sensor might be hacked by a strategic individual to manipulate the outcome to his/her own favour), or because users expect direct benefits from untrue reports (e.g., a retailer may wish to under-report local travel times on adjacent roads so as to mislead a routing application like Waze into diverting more traffic on a particular route, thus increasing its exposure).

Another context where data received from sensors may be strategically altered is when considering the cyber-security of distributed and/or networked systems. An important class of attacks for these systems is the so-called false data injection attack, whereby a malicious agent intercepts the original stream of measurements from sensors, and replaces it with corrupt data. 

While significant work has been devoted recently to characterizing the effects of corrupt data flows on closed-loop stability and devising identification procedures~\cite{6201214,cardenas2008research,mo2012cyber,Amin2013186,manshaei2011game,dan2010stealth,6545301}, relatively little has been done in the way of specifically modeling the attacker's strategic intent and understanding how the resulting system's behavior  differs from a mere failure mode. 

In this paper, we introduce and study a simple model of estimation in the presence of strategic, self-interested sensors. Specifically, we employ a game-theoretic setup to model the interaction between the sensor(s) and the receiver. The cost function of the receiver is taken to be equal to the variance of the estimation error of the state of nature while the cost function of the sensor has an extra term which is captured by its private information. As a starting point, we consider a static estimation problem in which a single sensor transmits a message about the state of nature in the presence of a (noisy but honest) side channel to the receiver. For this case, we show that there exists a family of simple equilibria (all resulting in the same estimation error variance). \farhad{At the captured equilibria, 
 the sensors \textit{never} ``flat-out lie'' and the receiver hence \textit{always} benefits from listening to the transmitted message.} We prove that, for some equilibria, the sensor's best response mapping does not utilize the side-channel information in constructing the message passed to the receiver (which, intuitively, makes sense as the receiver can always extract that part of the message since it also has access to the side-channel information). Using these results, we solve the dynamic counterpart of the proposed estimation problem. Interestingly, we prove that from the set of constructed equilibria for at least one equilibrium, the sensor employs a memory-less policy and, hence, the receiver uses a Kalman filter for constructing the state estimate. 

Equipped with these results, we extend the static estimation problem to multiple sensors with synchronous and asynchronous communication structures. First, we study synchronous communication structure, that is, the sensors transmit their messages simultaneously. Here, we restrict ourselves to the set of affine policies. \farhad{Although the assumption of affine policies for the sensors is rather restrictive,  it provides valuable insight because the provided analysis gives a lower-bound on the influence of the sensors (on the quality of the estimation), knowing that they find more degrees of freedom for constructing untruthful messages (and, hence, possibly creating a larger deviation to their benefit) with extending their set of available strategies to also cover nonlinear mappings.} Furthermore, we investigate symmetric problems in which the private information of the sensors are independently and identically distributed random variables.  We characterize an equilibrium of the game in this setup for which the quality of the receiver's estimation degrades as the number of sensors increases, which is a rather counter-intuitive result. We also investigate another notion of equilibrium, namely, the herding equilibrium, which supposes a lower degree of strategic behavior on the part of the sensors, and yields a very different scaling behavior for the estimation error. The herding scenario models an interesting intermediate situation where each sensor is refined enough to recognize that others may also be strategically misreporting but, having limited ``cognitive'' means or ability to predict the specific form of this behavior, assumes that they will just mimic its own action. Interestingly, for this equilibrium, the quality of the receiver's estimation improves as the number of sensors increases. Finally, we consider a multiple-sensor game under asynchronous communication structure, that is, the sensors transmit their messages sequentially. This is particularly useful if the sensors do not know the number of active participants in the estimation scheme (but they can observe, at least, the number of the sensor that have already contributed) as the the policy of each sensor, at the equilibrium, is only a function of the previous transmissions.

\farhad{The kind of self-interested strategic information transmission problems investigated in this paper has been considered before in the Economics literature, under the name of `cheap talk theory'. While specific assumptions (about the various priors' distributions and functional forms of the utility functions) vary, the basic framework for these problems (originally introduced in~\cite{Crawford1982}) involves two decision makers, a sender and a receiver, with different utility functions that both depend on a random state of nature and on the receiver's decision.  The sender's utility also depends on a parameter that is known solely to her and which is known in the literature as her ``private type".

The sender can directly observe the state of nature and decide which message (conditional on this observation) to transmit to the receiver. The receiver uses the message to modify his prior belief about the state of nature and, based on the new belief, makes a decision which impacts both utilities. The message itself does not \emph{directly} enter either utility, however, (it only matters to the extent that it modifies the receiver's belief distribution), which motivates the ``cheap talk" denomination. 

A central question in the cheap talk literature is the characterization of Nash equilibria, i.e., the determination of stochastic kernels for the sender and receiver which are best responses to each other. A fundamental result of~\cite{Crawford1982} (for situations where the state of nature is one-dimensional, compactly supported and uniformly distributed, and when the players' utilities are quadratic), is that the sender's strategy (mapping observation to transmitted signal) must employ quantization in every such equilibrium, with a computable upper-bound on the number of quantization cells. This can be interpreted by saying that strategic information transmission is parsimonious yet inevitably introduces confusion (due the non-injectivity of the sender's mapping).

Similar conclusions about the qualitative structure of Nash equilibria have since been derived in more general models of cheap talk, such as multidimensional sources~\cite{battaglini2002multiple}, noisy channels~\cite{yukselacc2015}, multiple senders~\cite{ambrus2008multi}, and hierarchical communication networks~\cite{ambrus2013hierarchical}. 

The model and the problem we consider in this paper differ from the traditional cheap talk framework detailed above in several significant ways. This brings this framework  closer to typical assumptions made in the controls literature, and is necessary to capture the motivating examples presented at the beginning of this introduction.  This also results in drastically different insights, since we show that, in our framework, various equilibria exist in which senders' and receiver's strategies are affine. More precisely, the differences are:
\begin{itemize}
\item[(1)] We assume that the state of nature is Gaussian with zero mean, when most of the cheap talk literature consider it to be compactly supported. This is particularly relevant for the data-attack example, where the role of ``state of nature" is played by the state of a dynamical system evolving under the action of some white process noise and, hence, takes value on the whole real line.
\item[(2)] The private type of the sender(s) is a random variable in our model, instead of a deterministic constant bounded away from zero. This is needed to capture situations where sensors do not know a priori by how much they will want to lie. This might occur, again, in the context of data attacks when the goal of the sensor is to manipulate reports so that the state estimated by the receiver tracks that of another legitimate-looking one (which is itself a stochastic process). Moreover, when considering a scenario with multiple sensors, the random generation of the private information ensures that we model various, often misaligned, incentives of a large pool of strategic sensors.
\item[(3)] Lastly but most importantly, we focus on Stackelberg equilibria rather than Nash equilibria (see, e.g.,~\cite{basar1999dynamic} as well as Section~\ref{sec:singlesensor} below for a rigorous definition of the former and some comparisons between the two). Although it has received relatively little attention in the cheap talk literature, this notion of equilibrium is more appropriate for the participatory sensing applications of interest to us, since the goal of the platform (which acts as the receiver) can legitimately be assumed to be known to the sensors (which act as senders). For example, users of Waze know that at least one goal of the system is to obtain an accurate estimate of travel times along all paths. It is thus justified to consider sensors as leaders, who act with the benefit of knowing that the receiver tries to minimize its estimation error, given their strategies.
\end{itemize}
}

\farhad{The problem of signalling in cooperative environments has attracted much attention dating back to the pioneering work of Witsenhausen in~\cite{Witsenhausen}. However, those studies do not assume any conflict of interest between the decision makers and, hence, the signalling enters the problem because of the trade-off between a precise communication and a perfect control using the same medium. Note that the communication is done through taking an action with potentially adverse effect on the control performance. This is evidently different from the presented framework in which the decision makers have a clear conflict of interest.}

The rest of the paper is organized as follows. First, we consider both static and dynamic estimation problems with a single sensor in the presence of a noisy but honest side-channel information in Section~\ref{sec:singlesensor}. In Sections~\ref{sec:multiplesensor}, we discuss the static multiple-sensor case under synchronous and asynchronous communication structures. Finally, we conclude the paper and present avenues for future research in Section~\ref{sec:conclusion}.

\subsection{Notation}
We shall let $\mathbb{R}$, $\mathbb{N}$, and $\mathbb{Z}$ denote the sets of real, natural, and integer numbers, respectively. Moreover, we define $\mathbb{N}_0=\mathbb{N}\cup \{0\}$. We use the notation $\N=\{n\in\mathbb{N}\,|\,n\leq N\}$. Furthermore, $\S_+^n$ and $\S_{++}^n$ denote the set of positive semi-definite and positive definite matrices in $\mathbb{R}^{n\times n}$. For any $A\in\mathbb{R}^{n\times n}$, we use the notations $A\geq 0$ and $A>0$ to denote $A\in \S_+^n$ and $A\in \S_{++}^n$, respectively. For any two random variables $x$ and $y$, we use the notation $\V_{xy}=\mathbb{E}\{xy^\top\}$. Let $\|x\|_2$ denote the 2-norm of vector $x\in\mathbb{R}^n$ for any $n\in\mathbb{N}$. For any matrix $A\in\mathbb{R}^{n\times n}$, $A^{\dag}$ is the Moore–Penrose pseudoinverse of $A$. Through out the paper, we also use the game theoretic convention $x=(x_i,x_{-i})$ in which $x_i$ and $x_{-i}$ denote $i$-th element of vector $x$ and the rest of its elements, respectively. Moreover, for time series, we define $x[k_1:k_2]=(x[k])_{k=k_1}^{k_2}$. \farhad{For any two arbitrary sets $\mathcal{X}$ and $\mathcal{Y}$, we define $\mathcal{C}(\mathcal{X},\mathcal{Y})$ to be the set of all Lebesgue-measurable mappings from $\mathcal{X}$ onto $\mathcal{Y}$.}

\section{Single Sensor with Side Information} \label{sec:singlesensor} 
In this section, we discuss static and dynamic state estimation with a single strategic sensor in the presence of an honest but noisy side-channel information.

\subsection{Static Estimation} \label{sec:singlesensor:static}
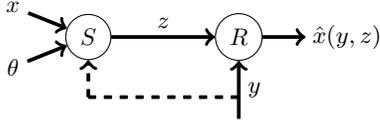
\begin{figure}[t]
\centering
\begin{tikzpicture}
\node[state,minimum size=0.1cm,scale=0.9] (R)  at (+1.0,+0.0) {$R$};
\node[state,minimum size=0.1cm,scale=0.9] (S)  at (-1.0,+0.0) {$S$};
\node[minimum size=0.01cm,scale=0.9] (I)  at (+1.0,-1.2) {};
\node[minimum size=0.01cm,scale=0.9] (x)  at (-2.0,+0.4) {$x$};
\node[minimum size=0.01cm,scale=0.9] (t)  at (-2.0,-0.4) {$\theta$};
\node[minimum size=0.01cm,scale=0.9] (h)  at (+2.0,-0.0) {};
\node[coordinate] (T1) at (+1.0,-0.8) {};
\node[coordinate] (T2) at (-1.0,-0.8) {};
\path[every node/.style={font=\sffamily\small}] 
			(S)  edge [->,double=black] node[above] {$z$} (R)
			(R)  edge [->,double=black] node[right] {\hspace{.1in}$\hat{x}(y,z)$} (h)
			(x)  edge [->,double=black] (S)
			(t)  edge [->,double=black] (S)
			(I)  edge [->,double=black] node[right] {$y$} (R)
			(T1) edge [- ,dashed,double=black] (T2)
			(T2) edge [->,dashed,double=black] (S);
\end{tikzpicture}
\vspace{-.1in}
\caption{\label{fig:diagram0} The communication structure between the strategic sensor $S$ and the side channel with the receiver $R$. We have used a dashed line to portray the availability of the side-channel information to the sensor because, as proved in the paper, there exists at least one equilibrium for which sensor does not use the (realization of the) side-channel information in constructing its message. }
\end{figure}

We consider the communication structure pictured in Figure~\ref{fig:diagram0}. The receiver (denoted by $R$ in Figure~\ref{fig:diagram0}) wants to estimate a random variable $x\in\mathbb{R}^{n_x}$. Throughout this subsection, we use the notation $\farhad{\upsilon(\cdot)}$ to denote this estimation as a function of the information available to the receiver. The sensor (denoted by $S$ in Figure~\ref{fig:diagram0}) transmits a signal $z\in\mathbb{R}^{n_z}$ (that may or may not contain some information about $x$). We assume that the sensor has access to the exact value of $x$. In addition to the message initiated by the sensor, the receiver also has access to a side channel that provides the measurement $y\in\mathbb{R}^{n_y}$. The timing of the game is as follows. First, the measurement $y$ is revealed. Then, sensor $S$ announces $z$. Finally, the receiver $R$ computes the optimal estimate by \farhad{minimizing $\mathbb{E}\{\|x-\farhad{\upsilon}(y,z)\|_2^2\}$ over $\Upsilon$ denoting} the set of all Lebesgue-measurable functions from $\mathbb{R}^{n_y}\times\mathbb{R}^{n_z}$ to $\mathbb{R}^{n_x}$. Let $\theta\in\mathbb{R}^{n_x}$ be the private information of the sensor (i.e., it is only available to the sensor~$S$). The sensor transmits a signal $z\in\mathbb{R}^{n_z}$ which is fully determined by the conditional distribution $p(\cdot|x,y,\theta)$. For the sake of brevity, and with abuse of notation, we refer to this as a ``stochastic mapping'' $z=\gamma(x,y,\theta)$ such that
$$
\mathbb{P}\{\gamma(x,y,\theta)\in\mathcal{Z}\}=\int_{z'\in\mathcal{Z}} p(z'|x,y,\theta) \mathrm{d}z', \,\, \forall \mathcal{Z}\subseteq \mathbb{R}^{n_z}.
$$
Let the set of all such mappings be denoted by $\Gamma$ (which has a one-to-one correspondence to the set of all the conditional distributions that construct the sensor's message). The goal of the sensor is to \farhad{minimize $\mathbb{E}\{\|(x+\theta)-\farhad{\upsilon}(y,\gamma(x,y,\theta))\|_2^2\}.$} Note that, currently, we assume that the sensor can access the side-channel information $y$ when constructing its message to the receiver. However, we will observe later that this assumption is not necessary, i.e., there exists at least one equilibrium for which the sensor does not utilize its knowledge of the side-channel information. \farhad{Hence, the receiver is trying to obtain the best estimate of $x$ (in the LMS sense) using the information available to him, while the sender, knowing that this is the goal of the receiver, chooses his message so as to  mislead the receiver in estimating $x+\theta$, where $\theta$ is a privately known parameter. Note that, for instance, in the case of  traffic estimation, the private information of the sensor is her desire, and its amount, for over-estimating or under-estimating the state of traffic on various links. This is certainly a private information as the other sensors and the estimator do not have access to it. The intention to miss-report the state of the traffic can be captured by that the sensor wants $\hat{x}=\upsilon(y,z)$, constructed by the received based the received measurements, to become equal to $x+\theta$ and, therefore, the sensor aims at minimizing the distance between these entities.
}

\farhad{
We need to define some useful notations before presenting the definition of the equilibrium of the game. For any $\hat{x}\in\mathcal{C}(\Gamma,\Upsilon)$ and any given $\gamma\in\Gamma$, $\hat{x}(\gamma)$ is a mapping in $\Upsilon$. We use the notation $[\hat{x}(\gamma)](y,z)$ to distinguish between the arguments of $\hat{x}$ and $\hat{x}(\gamma)$.

\begin{definition}\textsc{(Equilibrium):} \label{def:eq} A pair $(\hat{x}^*,\gamma^*)\in\mathcal{C}(\Gamma,\Upsilon)\times\Gamma$ constitutes an equilibrium if 
\begin{subequations}
\begin{align}
\hat{x}^* &\in\argmin_{\hat{x}\in\mathcal{C}(\Gamma,\Upsilon)} \mathbb{E}\{\|x-[\hat{x}(\gamma^*)](y,\gamma^*(x,y,\theta))\|_2^2\}, \\ 
\gamma^* &\in \argmin_{\gamma\in\Gamma} \mathbb{E}\{\|(x+\theta)-[\hat{x}^*(\gamma)](y,\gamma(x,y,\theta))\|_2^2\}.
\end{align}
\end{subequations}
\end{definition}

\begin{remark}[Stackelberg vs. Nash] Note that the equilibrium in Definition~\ref{def:eq} is not a Nash equilibrium but rather a Stackelberg equilibrium. This is because, in a Nash equilibrium, the players fix their policies; however, in this setup, the sensor explicitly calculates its best response assuming that the receiver is changing her estimation policy accordingly. 
%To brighten this aspect, we can define the equilibrium in an alternative fashion as follows. A pair $(\hat{x}^*,\gamma^*)\in\mathcal{C}(\Gamma,\Upsilon)\times\Gamma$ constitutes an equilibrium if, for any strategy $\gamma$ of the sender, the receiver employs
%\begin{align} \label{eqn:alt:def:1}
%\hat{x}^*(\gamma) &\in\argmin_{\upsilon\in\Upsilon} \mathbb{E}\{\|x-\upsilon(y,\gamma(x,y,\theta))\|_2^2\}
%\end{align}
%and the sender employs
%\begin{align}
%\gamma^* &\in \argmin_{\gamma\in\Gamma} \mathbb{E}\{\|(x+\theta)-[\hat{x}^*(\gamma)](y,\gamma(x,y,\theta))\|_2^2\}.
%\end{align}
%Here, we can clearly see that the receiver uses a best response policy for each $\gamma$ and, hence, her strategy changes in response to the strategy of the sender. Note that this alternative definition is well-defined if the set $\argmin_{\upsilon\in\Upsilon} \mathbb{E}\{\|x-\upsilon(y,\gamma(x,y,\theta))\|_2^2\}$ in~\eqref{eqn:alt:def:1} is a singleton which is the case if, for instance, the sender only uses affine policies for constructing its message.
\end{remark}

}

Throughout the rest of this subsection, we make the following assumption.

\begin{assumption} The random variables $x,y,\theta$ are jointly distributed Gaussian random variables with zero mean and a covariance matrix that satisfies
\begin{equation*}
\mathbb{E}\left\{ \matrix{c}{x\\ \theta\\ y} \matrix{c}{x\\ \theta\\ y}^\top \right\}
=\matrix{ccc}{\V_{xx} & \V_{x\theta} & \V_{x y} \\ \V_{\theta x} & \V_{\theta\theta} & \V_{\theta y} \\ \V_{yx} & \V_{y\theta} & \V_{yy}}\in\mathcal{S}_{++}^{2n_x+n_y}.
\end{equation*}
\end{assumption}

Contrary to the cheap-talk game literature~\cite{Crawford1982}, we assume that the private information of the sensor is a random variable which, as explained in the introduction, is relevant in situations of interest to this work. 

\begin{theorem} \label{tho:1}
There exists an equilibrium in which the receiver uses the Least Mean Square (LMS) estimator
\begin{equation} \label{eqn:recieverbestresponse}
\farhad{[\hat{x}^*(\gamma)]}(y,z)=\matrix{cc}{\V_{xy} & \V_{xz}} \matrix{cc}{\V_{yy} & \V_{yz} \\ \V_{zy} & \V_{zz}}^{-1}\matrix{c}{y\\ z},
\end{equation}
while the sensor uses the policy 
\begin{align} 
\gamma^*(x,y,\theta)=\alpha_1^\top x+\alpha_2^\top \theta+\alpha_3^\top y+v.\label{eqn:sensorbestresponse}
\end{align}
In the sender's policy, we have 
\begin{equation} \label{eqn:alpha}
\matrix{c}{\alpha_1 \\ \alpha_2 \\ \alpha_3}=\matrix{ccc}{\V_{xx} & \V_{x\theta} & \V_{x y} \\ \V_{\theta x} & \V_{\theta\theta} & \V_{\theta y} \\ \V_{yx} & \V_{y\theta} & \V_{yy}}^{-1}\matrix{c}{\V_{xz} \\ \V_{\theta z} \\ \V_{yz}},
\end{equation}
and $v\in\mathbb{R}^{n_z}$ is a Gaussian random variable with zero mean and covariance matrices
\begin{subequations} \label{eqn:covariancevvvv}
\begin{equation} \label{eqn:covariancev}
\matrix{ccc}{\V_{vx} & \V_{v\theta} & \V_{vy}}=\matrix{ccc}{0 & 0 &0 },
\end{equation}
\begin{equation} \label{eqn:covariancevv}
V_{vv}=I-\matrix{c}{ \V_{xz} \\ \V_{\theta z} \\ \V_{yz}}^\top Q \matrix{c}{ \V_{xz} \\ \V_{\theta z} \\ \V_{yz}},
\end{equation}
\end{subequations}
where
\begin{subequations}\label{eqn:tho1:optimization}
\begin{align} 
\matrix{c}{ \V_{xz} \\ \V_{\theta z} \\ \V_{yz}}\in\mathcal{X}=\argmin_{\xi\in\mathbb{R}^{(2n_x+n_y)\times n_z}} & \trace\left(\xi^\top W \xi\right),\\
\mathrm{s.t.} \hspace{.3in}& \,\,\xi^\top Q \xi\leq I,
\end{align}
\end{subequations}
with 
\begin{subequations} \label{eqn:definition:WQ}
\begin{align} 
W=\Xi^\top\matrix{cc}{-I & -I \\ -I & 0}\Xi,\label{eqn:definition:W}
\end{align}
\begin{equation} \label{eqn:definition:Q}
\begin{split}
Q&=\Xi^\top\hspace{-.04in}
\left(\hspace{-.04in}\matrix{ccc}{\V_{xx} & \V_{x\theta} \\ \V_{\theta x} & \V_{\theta\theta}}\hspace{-.04in}-\hspace{-.04in}\matrix{c}{\V_{xy} \\ \V_{\theta y}}\hspace{-.03in}\V_{yy}^{-1}\hspace{-.03in} \matrix{c}{\V_{xy} \\ \V_{\theta y}}^{\hspace{-.04in}\top}\right)^{\hspace{-.04in}-1}\hspace{-.04in}\Xi,
\end{split}
\end{equation}
\begin{align} \label{eqn:definition:Xi}
\Xi=\matrix{ccc}{I & 0 & -\V_{xy}\V_{yy}^{-1}\\ 0 & I & -\V_{\theta y}\V_{yy}^{-1}}.
\end{align}
\end{subequations}
Furthermore, the sensor's policy of the form $\kappa\gamma^*$, for some $\kappa\in \mathbb{R}\setminus\{0\}$, along side the receiver's policy $\hat{x}^*$, also constitutes an equilibrium. All these equilibria result in the same estimation error variance at the receiver.
\end{theorem}

\begin{IEEEproof} When the sensor uses the strategy in~\eqref{eqn:sensorbestresponse}, the receiver's best response is the LMS estimator~\cite[p.\,80]{kailath2000linear}. We thus only need to show that, provided the receiver uses \farhad{$[\hat{x}^*(\gamma^*)](\cdot)$} in~\eqref{eqn:recieverbestresponse}, the sensor's optimal policy is indeed linear, and satisfies~\eqref{eqn:alpha}--\eqref{eqn:definition:WQ}. Note that, once the receiver's strategy is fixed as above, the sensor's cost can be written solely as a function of $V_{zx}$, $V_{zy}$, and $V_{z\theta}$. Notice that $V_{zx}=\mathbb{E}\{z x^\top\}$, $V_{zy}=\mathbb{E}\{z y^\top\}$, and $V_{z\theta}=\mathbb{E}\{z \theta^\top\}$ are not mere constants but they are functions of the policy of the sender $\gamma\in\Gamma$. Indeed, we can write~\eqref{eqn:long:proof1}\farhad{, which is presented on top of the next page}. 
\begin{figure*}
\begin{align}
\mathbb{E}\{\|(x+\theta)-\hat{x}^*(y,z)\|_2^2\}
%&=\trace (\mathbb{E}\{((x+\theta)-\hat{x}^*(y,z))((x+\theta)-\hat{x}^*(y,z))^\top\}) \nonumber \\
&=\trace\bigg(\V_{xx}+\V_{x\theta}-\matrix{cc}{\V_{xy} & \V_{xz}} \matrix{cc}{\V_{yy} & \V_{yz} \\ \V_{zy} & \V_{zz}}^{-1} \matrix{c}{\V_{yx} \\ \V_{zx}}+\V_{\theta x}+
\V_{\theta\theta}\nonumber\\&\hspace{.2in} -\hspace{-.04in}\matrix{cc}{\V_{\theta y} & \V_{\theta z}} \hspace{-.04in}\matrix{cc}{\V_{yy} & \V_{yz} \\ \V_{zy} & \V_{zz}}^{-1} \matrix{c}{\V_{yx} \\ \V_{zx}} \hspace{-.04in}-\hspace{-.04in}\matrix{cc}{\V_{xy} & \V_{xz}} \matrix{cc}{\V_{yy} & \V_{yz} \\ \V_{zy} & \V_{zz}}^{-1}\hspace{-.04in}\matrix{c}{\V_{y\theta}\\ \V_{z\theta}} \bigg). \label{eqn:long:proof1}
\end{align}
\hrulefill
\end{figure*}
In the rest of the proof, without loss of generality, we assume $\V_{zz}-\V_{zy}\V_{yy}^{-1}\V_{yz}=I$. Note that this is without loss of generality because the optimization problem after scaling $z$ is feasible if and only if it is feasible before scaling $z$. Following Item~1, Section~3.5.3 in~\cite[pp.\,29-30]{lkepohl1996handbook}, gives
\begin{equation*}
\begin{split}
\matrix{cc}{\V_{yy} & \V_{yz} \\ \V_{zy} & \V_{zz}}^{-1}&\hspace{-.06in}
%=\hspace{-.05in}\matrix{cc}{I & -\V_{yy}^{-1}\V_{yz} \\ 0 & I} \\ &\hspace{.2in}\times\matrix{cc}{\V_{yy}^{-1} & 0 \\ 0 & (\V_{zz}-\V_{zy}\V_{yy}^{-1}\V_{yz})^{-1}} \\ &\hspace{.2in}\times\matrix{cc}{I & 0 \\ -\V_{zy}\V_{yy}^{-1} & I}
%\\
\hspace{-.06in}=\hspace{-.05in} \matrix{cc}{\V_{yy}^{-1}+\V_{yy}^{-1}\V_{yz}\V_{zy}\V_{yy}^{-1} & -\V_{yy}^{-1}\V_{yz} \\ -\V_{zy}\V_{yy}^{-1} & I}\hspace{-.03in}.
\end{split}
\end{equation*}
Substituting this identity into~\eqref{eqn:long:proof1}, we can observe that
\begin{equation*}
\begin{split}
\mathbb{E}&\{\|(x+\theta)-\hat{x}\|_2^2\}=\\
%=\trace\big(-
%\V_{zy}\V_{yy}^{-1}\V_{yx}\V_{xy}\V_{yy}^{-1}\V_{yz}\\&+
%\V_{zx}\V_{xy}\V_{yy}^{-1}\V_{yz}+
%\V_{zy}\V_{yy}^{-1}\V_{yx}\V_{xz}-
%\V_{zx}\V_{xz}
%\\&-
%\V_{zy}\V_{yy}^{-1}\V_{yx}\V_{\theta y}\V_{yy}^{-1}\V_{yz}+
%\V_{zx}\V_{\theta y}\V_{yy}^{-1}\V_{yz}\\&+
%\V_{zy}\V_{yy}^{-1}\V_{yx}\V_{\theta z}-
%\V_{zx}\V_{\theta z}-
%\V_{zy}\V_{yy}^{-1}\V_{y\theta}\V_{xy}\V_{yy}^{-1}\V_{yz}\\&+
%\V_{z\theta}\V_{xy}\V_{yy}^{-1}\V_{yz}+
%\V_{zy}\V_{yy}^{-1}\V_{y\theta}\V_{xz}-
%\V_{z\theta}\V_{xz}
%\big)+c\\
&=\trace\left(
\matrix{ccc}{\V_{zx} &  \V_{z\theta} &  \V_{zy}} W
\matrix{c}{\V_{xz} \\ \V_{\theta z} \\ \V_{yz}}
\right)+c,
\end{split}
\end{equation*}
where $W$ is defined in~\eqref{eqn:definition:W} and 
$c=\trace(\V_{xx}+\V_{x\theta}+\V_{\theta x}+\V_{\theta\theta}-\V_{xy}\V_{yy}^{-1}\V_{yx}-\V_{\theta y}\V_{yy}^{-1}\V_{yx}-\V_{xy}\V_{yy}^{-1}\V_{y\theta})$
does not depend on the sensor's strategy. Since the covariance matrix of the vector of random variables  $[x^\top \, \theta^\top \, y^\top \, z^\top]^\top$ is a positive semi-definite matrix, it should satisfy 
\begin{equation}\label{eqn:proof:positive_semi-definite_matrix}
\mathbb{E}\left\{ \matrix{c}{x\\ \theta\\ y \\ z} \matrix{c}{x\\ \theta\\ y \\ z}^\top \right\}
=\matrix{cccc}{\V_{xx} & \V_{x\theta} & \V_{xy} & \V_{xz}\\ \V_{\theta x} & \V_{\theta\theta} & \V_{\theta y} & \V_{\theta z} \\ \V_{yx} & \V_{y\theta} & \V_{yy} & \V_{yz} \\ \V_{zx} & \V_{z\theta} & \V_{zy} & \V_{zz}} \geq 0. 
\end{equation}
Note that $\V_{zz}=\V_{zy}\V_{yy}^{-1}\V_{yz}+I>0$. We use the Schur complement 
%(see~\cite[p.\,651]{boyd2004convex}) 
to show that the condition in~\eqref{eqn:proof:positive_semi-definite_matrix} is equivalent to
%\begin{equation*}
%\begin{split}
%&-
%\matrix{c}{ \V_{xz}\\ \V_{\theta z} \\ \V_{yz}}^\top
%\matrix{ccc}{\V_{xx} & \V_{x\theta} & \V_{xy} \\ \V_{\theta x} & \V_{\theta\theta} & \V_{\theta y} \\ \V_{yx} & \V_{y\theta} & \V_{yy}}^{-1}
%\matrix{c}{ \V_{xz}\\ \V_{\theta z} \\ \V_{yz}}\\ &\hspace{1.9in}+ I+\V_{zy}\V_{yy}^{-1}\V_{yz}\geq 0,
%\end{split}
%\end{equation*}
%which is, in turn, equivalent to
\begin{equation*}
\begin{split}
&\matrix{c}{ \V_{xz}\\ \V_{\theta z} \\ \V_{yz}}^\top
\left(\matrix{ccc}{\V_{xx} & \V_{x\theta} & \V_{xy} \\ \V_{\theta x} & \V_{\theta\theta} & \V_{\theta y} \\ \V_{yx} & \V_{y\theta} & \V_{yy}}^{-1}\right. \\ &\hspace{1.in}\left.-\matrix{ccc}{0 & 0 & 0 \\ 0 & 0 & 0 \\ 0 & 0 & \V_{yy}^{-1}}\right)
\matrix{c}{ \V_{xz}\\ \V_{\theta z} \\ \V_{yz}}\leq I.
\end{split}
\end{equation*}
Now, using Item~(2) in Section~3.5.3 in~\cite[p.\,30]{lkepohl1996handbook}, we can easily prove the identity in~\eqref{eqn:derivation:Q}\farhad{, on top of the next page,} in which 
\begin{figure*}
\begin{align} 
\matrix{ccc}{\hspace{-.05in}\V_{xx} & \hspace{-.05in}\V_{x\theta} \hspace{-.05in}& \V_{xy} \hspace{-.05in}\\ \hspace{-.05in}\V_{\theta x} & \hspace{-.05in}\V_{\theta\theta}\hspace{-.05in} & \V_{\theta y} \hspace{-.05in}\\ \hspace{-.05in}\V_{yx} & \hspace{-.05in}\V_{y\theta}\hspace{-.05in} & \V_{yy}\hspace{-.05in}}^{\hspace{-.04in}-1}\hspace{-.07in}-\hspace{-.05in}\matrix{ccc}{0 & 0 & 0 \\ 0 & 0 & 0 \\ 0 & 0 & \V_{yy}^{-1}}&\hspace{-.05in}=\hspace{-.05in}
\matrix{c;{2pt/2pt}c}{ J & \hspace{-.09in}-J\matrix{c}{\V_{xy} \\ \V_{\theta y}}\V_{yy}^{-1} \\[3mm] \hdashline \hspace{-.09in}-\V_{yy}^{-1}\hspace{-.05in}\matrix{cc}{\V_{yx} & \V_{y\theta}} \hspace{-.05in}J\hspace{-.05in} & \hspace{-.05in}\V_{yy}^{-1}\hspace{-.05in}\matrix{cc}{\V_{yx} & \V_{y\theta}}\hspace{-.03in} J\hspace{-.03in} \matrix{c}{\V_{xy} \\ \V_{\theta y}}\hspace{-.05in}\V_{yy}^{-1}\hspace{-.03in}+\hspace{-.03in}\V_{yy}^{-1} \hspace{-.08in}}\hspace{-.07in}-\hspace{-.05in}
\matrix{c;{2pt/2pt}c}{ 0 & 0 \\ \hdashline 0 & \V_{yy}^{-1} }\nonumber\\[.4em]
&\hspace{-.05in}=\hspace{-.05in}
\matrix{c;{2pt/2pt}c}{ J & \hspace{-.09in}-J\matrix{c}{\V_{xy} \\ \V_{\theta y}}\V_{yy}^{-1} \\[3mm] \hdashline \hspace{-.09in}-\V_{yy}^{-1}\hspace{-.05in}\matrix{cc}{\V_{yx} & \V_{y\theta}} \hspace{-.05in}J\hspace{-.05in} & \hspace{-.05in}\V_{yy}^{-1}\hspace{-.05in}\matrix{cc}{\V_{yx} & \V_{y\theta}}\hspace{-.03in} J\hspace{-.03in} \matrix{c}{\V_{xy} \\ \V_{\theta y}}\hspace{-.05in}\V_{yy}^{-1}}\nonumber\\
&\hspace{-.05in}=\hspace{-.05in}
\matrix{ccc}{I & 0 & -\V_{xy}\V_{yy}^{-1}\\ 0 & I & -\V_{\theta y}\V_{yy}^{-1}}^\top
J\matrix{ccc}{I & 0 & -\V_{xy}\V_{yy}^{-1}\\ 0 & I & -\V_{\theta y}\V_{yy}^{-1}}.\label{eqn:derivation:Q}
\end{align}
\hrulefill
\end{figure*}
\begin{equation} \label{eqn:proof:def:J}
J=\left(\matrix{ccc}{\V_{xx} & \V_{x\theta} \\ \V_{\theta x} & \V_{\theta\theta}}-\matrix{c}{\V_{xy} \\ \V_{\theta y}}\V_{yy}^{-1} \matrix{c}{\V_{xy} \\ \V_{\theta y}}^\top\right)^{-1}.
\end{equation}
Therefore, the sensor's best response can be extracted from solving the optimization problem in~\eqref{eqn:tho1:optimization}. Now, we just need to show that there exists an affine policy for the sensor that results in covariance matrices $\V_{zx},\V_{z\theta},\V_{zy}$. Let $z=\alpha_1^\top x+\alpha_2^\top \theta+\alpha_3^\top y+v$ where $v\in\mathbb{R}^{n_z}$ is a Gaussian random variable. In this case, we can compute
\begin{align*}
\matrix{c}{\V_{xz} \\ \V_{\theta z} \\ \V_{yz}}
%&=\matrix{c}{\V_{xx}\alpha_1+\V_{x\theta}\alpha_2+\V_{xy}\alpha_3 \\ \V_{\theta x}\alpha_1+\V_{\theta\theta}\alpha_2+\V_{\theta y}\alpha_3 \\ \V_{yx}\alpha_1+\V_{y\theta}\alpha_2+\V_{yy}\alpha_3 }+\matrix{c}{\V_{xv} \\ \V_{\theta v} \\ \V_{yv}}\\&
=\matrix{ccc}{\V_{xx} & \V_{x\theta} & \V_{x y} \\ \V_{\theta x} & \V_{\theta\theta} & \V_{\theta y} \\ \V_{yx} & \V_{y\theta} & \V_{yy}} \matrix{c}{\alpha_1 \\ \alpha_2 \\ \alpha_3}+\matrix{c}{\V_{xv} \\ \V_{\theta v} \\ \V_{yv}},
\end{align*}
which results in~\eqref{eqn:alpha} and~\eqref{eqn:covariancev}. Moreover,~\eqref{eqn:covariancevv} follows from substituting~\eqref{eqn:alpha} and~\eqref{eqn:covariancev} into $\V_{zz}-\V_{zy}\V_{yy}^{-1}\V_{yz}=I$.
\end{IEEEproof}

\begin{remark} In~\eqref{eqn:recieverbestresponse}, the policy of the receiver is affine in the realization of the messages $y,z$. However, overall, this policy is not affine as $\V_{xz},\V_{yz},\V_{zz}$ are all functions of the random variable $z$'s distribution (but not its realization). \farhad{The dependency of the gains to these covariance matrices clearly illustrates the dependency of $\hat{x}^*\in\mathcal{C}(\Gamma,\Upsilon)$ to $\gamma\in\Gamma$.}
\end{remark}

\begin{remark} Note that even when $x$ and $\theta$ are uncorrelated, the sender always sends ``some amount of information about $x$'' rather than just sending $\theta$. This is indeed true because if $z$ does not contain any information about the state of nature, the receiver will simply discard it (if $y$ is not correlated with $\theta$ and if there is such a correlation, the receiver uses $z$ to cancel out the correlation). Hence, it is always in the receiver's best interest to listen to any sensor, be it strategic or not. \farhad{We formalize this observation for the special case of scalar message later in Proposition~\ref{prop:1}.}
\end{remark} 

Note that the optimization problem in~\eqref{eqn:tho1:optimization} is not a convex optimization problem as the matrix $W$ is an indefinite matrix. Therefore, solving it numerically is, in general, a tedious task and it would be of interest to find an explicit solution, at least, under some conditions. One such case is discussed in the following corollary, which considers the case where the sensor's message is scalar.

\begin{corollary} \label{cor:1} Let $n_z=1$. There exists an equilibrium in which the receiver uses the LMS estimator in~\eqref{eqn:recieverbestresponse} while the sensor uses the policy in~\eqref{eqn:sensorbestresponse}. In the sender's policy, $\alpha_1,\alpha_2,\alpha_3$ are defined using~\eqref{eqn:alpha} and $v\in\mathbb{R}^{n_z}$ is a Gaussian random variable with zero mean and covariance matrices as in~\eqref{eqn:covariancevvvv}, where $\V_{zx},\V_{z\theta},\V_{zy}$ are determined by
\begin{equation*}
\begin{split}
\matrix{c}{ \V_{xz} \\ \V_{\theta z} \\ \V_{yz}}=\matrix{ccc}{I & 0 & -\V_{xy}\V_{yy}^{-1}\\ 0 & I & -\V_{\theta y}\V_{yy}^{-1}}^{\dag}J^{-1/2}\pi,
\end{split}
\end{equation*}
and $\pi$ denotes the normalized eigenvector (i.e. $\|\pi\|_2=1$) of the smallest eigenvalue of 
\begin{equation*}
\begin{split}
E=J^{-1/2}\matrix{cc}{-I & -I \\ -I & 0} J^{-1/2}
\end{split}
\end{equation*}
with $J$ defined as in~\eqref{eqn:proof:def:J}. Furthermore, the sensor's policy of the form $\kappa\gamma^*$, for some $\kappa\in \mathbb{R}\setminus\{0\}$, along side the receiver's policy $\hat{x}^*$, also constitutes an equilibrium.
\end{corollary}

\begin{IEEEproof} With the change of variable
\begin{equation*}
\begin{split}
\eta
&=\matrix{ccc}{I & 0 & -\V_{xy}\V_{yy}^{-1}\\ 0 & I & -\V_{\theta y}\V_{yy}^{-1}}\matrix{c}{ \V_{xz} \\ \V_{\theta z} \\ \V_{yz}},
\end{split}
\end{equation*}
we can rewrite the optimization problem in~\eqref{eqn:tho1:optimization} as
\begin{equation} \label{eqn:cor1:optimization}
\begin{split}
\min_{\eta\in\mathbb{R}^{2n_x}} \,\,&  \eta^\top \matrix{cc}{-I & -I \\ -I & 0} \eta,\\
\mathrm{s.t.} \,\,\,\,\,& \eta^\top J \eta\leq 1.
\end{split}
\end{equation}
Now, letting $\bar{\eta}=J^{1/2}\eta$ results in
\begin{equation} \label{eqn:cor1:optimization:2}
\begin{split}
\min_{\bar{\eta}\in\mathbb{R}^{2n_x}} \,\,& \bar{\eta}^\top J^{-1/2}\matrix{cc}{-I & -I \\ -I & 0} J^{-1/2}\bar{\eta},\\
\mathrm{s.t.} \,\,\,\,\,& \bar{\eta}^\top \bar{\eta}\leq 1.
\end{split}
\end{equation}
Notice that the matrix $E$, which appears in the cost function of~\eqref{eqn:cor1:optimization:2}, has, at least, one negative eigenvalue. This is because multiplying a matrix from both sides by a symmetric and invertible matrix does not change the sign of its eigenvalues (see Sylvester's law of inertia~\cite[p.\,282]{horn2012matrix}). Therefore, using Lemma~\ref{lemma:eigenvector} in Appendix~\ref{App:UsefulLemma}, we realize that the solution to the optimization problem in~\eqref{eqn:cor1:optimization:2} is the normalized eigenvector corresponding to the smallest eigenvalue of $E$.
\end{IEEEproof}

Before moving on to the dynamic estimation problem, we show that there exists at least \farhad{one} equilibrium for which the sender's best response does not depend on the side-channel information. This is of special interest to us because, typically, the side-channel information might be encrypted (and, hence, not accessible to the sensor) or the sensor and the receiver might not be co-located (and, hence, the sensor might not have the opportunity to eavesdrop on this information). Intuitively, such an equilibrium exists because the receiver can always construct $\tilde{z}=z-Ky$, for some $K\in\mathbb{R}^{n_z\times n_y}$ such that $\tilde{z}$ and $y$ are uncorrelated, with the same amount of information content because $\spans(z,y)=\spans(\tilde{z},y)$. This result also provides the opportunity to extend the framework to the case where the side channel and the strategic sensor reveal their messages simultaneously. Note that the following corollary holds for all $n_z\geq 1$.

\begin{corollary} \label{cor:notafunctionofy}
There exists an equilibrium in which $\gamma^*$ is independent of $y$. More precisely, the receiver uses the LMS estimator in~\eqref{eqn:recieverbestresponse} while the sensor uses the policy 
\begin{align} \label{eqn:sensorbestresponse:1}
\gamma^*(x,y,\theta)=\alpha_1^\top x+\alpha_2^\top \theta+v,
\end{align}
where $v\in\mathbb{R}^{n_z}$ is a Gaussian random variable with zero mean and covariance matrices
\begin{subequations} \label{eqn:covariancevvvv:1}
\begin{equation}
\matrix{ccc}{\V_{vx} & \V_{v\theta} & \V_{vy}}=\matrix{ccc}{0 & 0 &0 },
\end{equation}
\begin{equation} 
V_{vv}=I-\matrix{c}{\alpha_1 \\ \alpha_2}^\top \Xi' \matrix{c}{\alpha_1 \\ \alpha_2},
\end{equation}
\end{subequations}
in which 
\begin{subequations} \label{eqn:cor:notafunctionofy:optimzation}
\begin{align} 
\matrix{c}{\alpha_1 \\ \alpha_2}\in\mathcal{X}'=\argmin_{\xi\in\mathbb{R}^{2n_x\times n_z}} & \trace\left({\xi'}^\top W' \xi'\right),\\
\mathrm{s.t.} \hspace{.18in}& \,\,{\xi'}^\top \Xi' {\xi'}\leq I,
\end{align}
\end{subequations}
with 
\begin{subequations} 
\begin{align} 
W'={\Xi'}^\top\matrix{cc}{-I & -I \\ -I & 0}{\Xi'},
\end{align}
\begin{align}
\Xi'=\matrix{ccc}{\V_{xx} & \V_{x\theta} \\ \V_{\theta x} & \V_{\theta\theta}}-\matrix{c}{\V_{xy} \\ \V_{\theta y}}\V_{yy}^{-1}\matrix{c}{\V_{xy} \\ \V_{\theta y}}^{\top}.
\end{align}
\end{subequations}
Furthermore, the sensor's policy of the form $\kappa\gamma^*$, for some $\kappa\in \mathbb{R}\setminus\{0\}$, along side the receiver's policy $\hat{x}^*$, also constitutes an equilibrium. All these equilibria result in the same estimation error variance at the receiver.
\end{corollary}

\begin{IEEEproof} Let us introduce the change of variable
$$
\xi=\matrix{ccc}{\V_{xx} & \V_{x\theta} & \V_{x y} \\ \V_{\theta x} & \V_{\theta\theta} & \V_{\theta y} \\ \V_{yx} & \V_{y\theta} & \V_{yy}}\tilde{\xi}.
$$
\farhad{Recalling the definition of $\Xi$ in~\eqref{eqn:definition:Xi}, we get}
\begin{align*}
\Xi\xi
&=\matrix{ccc}{
\V_{xx}-\V_{xy}\V_{yy}^{-1}\V_{yx} & \V_{x\theta}-\V_{xy}\V_{yy}^{-1}\V_{y\theta} & 0\\ 
\V_{\theta x}-\V_{\theta y}\V_{yy}^{-1}\V_{yx} & \V_{\theta\theta}-\V_{\theta y}\V_{yy}^{-1}\V_{y\theta} & 0}\tilde{\xi}\\
&=\Xi'\matrix{cc}{I_{2n_x} & 0_{2n_x\times n_y}}\tilde{\xi}\\
&=\Xi'\xi',
\end{align*}
where $\xi'=\matrix{cc}{I_{2n_x} & 0_{2n_x\times n_y}}\tilde{\xi}$. As a result, 
\begin{align*}
\xi^\top Q\xi&={\xi'}^\top \Xi'{\xi'},\\
\xi^\top W\xi&={\xi'}^\top \Xi' \matrix{cc}{-I & -I \\ -I & 0}\Xi'{\xi'}={\xi'}^\top W'{\xi'},
\end{align*}
Therefore, this change of variable transforms the optimization problem in~\eqref{eqn:tho1:optimization} to the one in~\eqref{eqn:cor:notafunctionofy:optimzation} and, as a result, we get
\begin{align*}
\left\{\matrix{ccc}{\V_{xx} & \V_{x\theta} & \V_{x y} \\ \V_{\theta x} & \V_{\theta\theta} & \V_{\theta y} \\ \V_{yx} & \V_{y\theta} & \V_{yy}}\matrix{c}{\xi' \\ 0_{n_y\times n_z}} \bigg| \xi'\in\mathcal{X}' \right\}\subseteq \mathcal{X}.
\end{align*}
Therefore, from~\eqref{eqn:alpha}, there exists an equilibrium for which
$$
\matrix{c}{\alpha_1 \\ \alpha_2 \\ \alpha_3}=\matrix{c}{\xi' \\ 0_{n_y\times n_z}},
$$
where $\xi'\in\mathcal{X}'$. For this equilibrium, clearly, $\alpha_3=0$. 
\end{IEEEproof}

\farhad{
Signalling games, such as cheap-talk games, most often admit a family of trivial equilibria, known as babbling equilibria~\cite{Sobel2009Chapter}, in which (\textit{i}) the transmitted signal of the sensor is not correlated with the to-be-estimated random variable and (\textit{ii}) the receiver completely dismisses the transmitted signal of the sensor. In what follows, we prove that the equilibrium captured in Corollary~\ref{cor:1} is not a babbling one, i.e., the sensor does not ``flat-out lie'' and the receiver hence benefits from listening to the transmitted message. Moreover, at the recovered equilibrium, complete honesty is never in the sensor's benefit.

\begin{proposition} \label{prop:1} Let $n_z=1$. For the equilibria captured in Corollary~\ref{cor:notafunctionofy}, we have $\alpha_1\neq 0$ and $\alpha_2\neq 0$.
\end{proposition}

\begin{IEEEproof} \farhad{See Appendix~\ref{proof:prop:1}.}
%Please see~\cite{FarokhiCompleteManuscript2015} for a detailed proof.
\end{IEEEproof}

Clearly, because $\alpha_1\neq 0$, at the equilibrium, the sensor's message always carries some useful information. Moreover, the message also partially reflects the private information of the sensor because $\alpha_2\neq 0$. Notice that this result holds irrespective of the correlation between $x$ and $\theta$. Let us show this with help of a small example.

\begin{figure}
\centering
\includegraphics[width=0.9\linewidth]{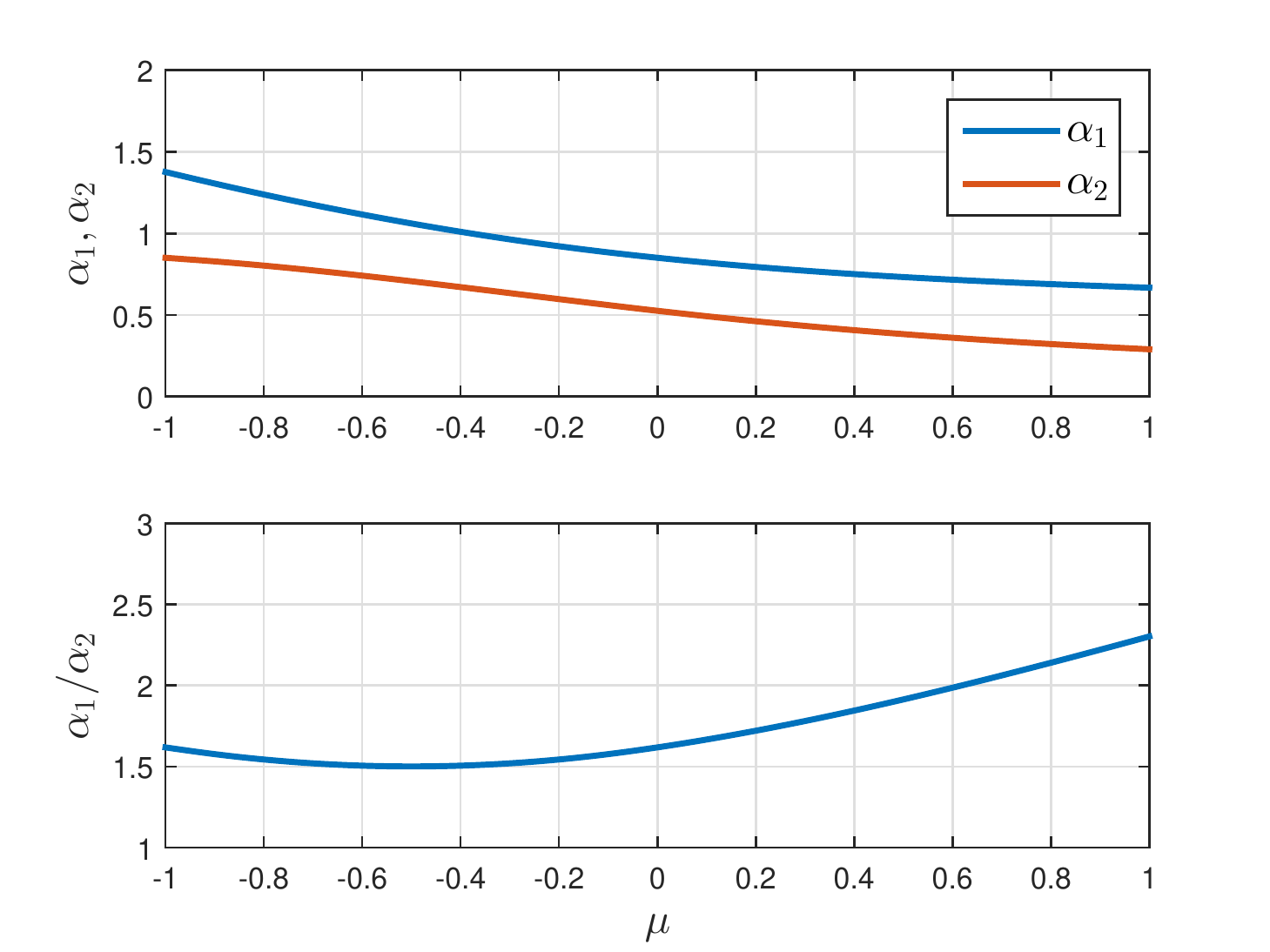}
\caption{\label{fig:variation} Gains $\alpha_1,\alpha_2$ and their ratio at the equilibrium as a function of $\mu$. }
\end{figure}

\begin{example} Let us consider the simple setup in which $V_{xx}=1$ and there is no side channel information available. Further, assume that $\theta=\mu x+n$ where $n$ is an independent Gaussian random variable with zero mean and $V_{nn}=1$. This way, we can capture a variety of interesting cases. Doing so, we get $\V_{x\theta}=\mu$ and $V_{\theta\theta}=\mu^2+1$. Figure~\ref{fig:variation} illustrates coefficients $\alpha_1,\alpha_2$ and their ratio, at the captured equilibrium, as a function of the correlation between $x$ and $\theta$. Interestingly, even for negative correlations, the message contains some useful information about $x$. For instance, even when $\mu=-1$ which points to that the sensor and the receiver have completely different objectives, the sensor's message contains significant information about the variable $x$. The same is true for the $\mu=0$, pointing to the case where there is no correlation between $\theta$ and $x$.

For the setup of this example, we can calculate the equilibrium explicitly as
\begin{align*}
\begin{bmatrix}
\alpha_1 \\ \alpha_2
\end{bmatrix}=&\frac{1}{\sqrt{\big(2\mu + \sqrt{(2\mu+1)^2 + 4} + 1\big)^2 + 4}}\\&\hspace{1.2in}\times 
\begin{bmatrix}
\sqrt{(2\mu+1)^2 + 4} + 1\\
2
\end{bmatrix}.
\end{align*}
Further, we have
\begin{align*}
\frac{\alpha_1}{\alpha_2}=\frac{\sqrt{(2\mu+1)^2 + 4} + 1}{2}.
\end{align*}
In this case, we have
\begin{align*}
\frac{\mathrm{d}}{\mathrm{d}\mu} \bigg(\frac{\alpha_1}{\alpha_2} \bigg)
=\frac{2\mu+1}{\sqrt{(2\mu+1)^2+4}}.
\end{align*}
Evidently, after the threshold $\mu>-1/2$, as we increase the correlation, the sensor provides a more accurate measurement of the to-be-estimated random variable $x$ (because the ratio $\alpha_1/\alpha_2$ becomes an increasing function of $\mu$).
%First, we consider the case where there is no correlation between $x$ and $\theta$, i.e., $V_{x\theta}=0$. For this case, at the equilibrium, we have
%\begin{align*}
%\matrix{c}{\alpha_1 \\ \alpha_2}=\matrix{c}{0.8507
% \\  0.5257}.
%\end{align*}
%As we expect $\alpha_1\neq 0$ and, therefore, the transmitted signal contains some useful information about the realization of $x$. 
%
%Figure~\ref{fig:variation} illustrates coefficients $\alpha_1,\alpha_2$ at the captured equilibrium as a function of the correlation between $x$ and $\theta$. Evidently, as we increase the correlation, the sensor provides a more accurate measurement of the to-be-estimated random variable $x$. Interestingly, even for negative correlations, the message contains some useful information about $x$.
%
%Now, consider the case where $\theta=-x+n$ where $n$ is a Gaussian random variable with zero mean and $V_{nn}=1$. Recalling the cost of the sensor, it is easy to see that, in this case, the sensor and the receiver have completely different objectives. We have $V_{x\theta}=-V_{xx}$ and $V_{\theta\theta}=V_{xx}+V_{nn}=2$. In this case, at the equilibrium, we get
%\begin{align*}
%\matrix{c}{\alpha_1 \\ \alpha_2}=\matrix{c}{1.3764
% \\ 0.8507}.
%\end{align*}
%Even in the extreme case that the sensor and the receiver's interests are completely mismatched, the sensor provides some useful information regarding the random variable $x$ because, otherwise, the receiver can simply ignore the transmitted message which is worse for the sensor. 
\end{example} 
}

Now, let us extend the presented formulation to dynamic estimation problem.

\subsection{Dynamic Estimation} \label{sec:singlesensor:dynamic}
\begin{figure}[t]
\centering
\begin{tikzpicture}
\node[state,minimum size=0.1cm,scale=0.9] (R)  at (+1.0,+0.0) {$R$};
\node[state,minimum size=0.1cm,scale=0.9] (S)  at (-1.0,+0.0) {$S$};
\node[minimum size=0.01cm,scale=0.9] (I)  at (+1.0,-1.2) {};
\node[minimum size=0.01cm,scale=0.9] (x)  at (-2.0,+0.4) {$x[k]$};
\node[minimum size=0.01cm,scale=0.9] (t)  at (-2.0,-0.4) {$\theta[k]$};
\node[minimum size=0.01cm,scale=0.9] (h)  at (+2.0,-0.0) {};
\node[coordinate] (T1) at (+1.0,-0.8) {};
\node[coordinate] (T2) at (-1.0,-0.8) {};
\path[every node/.style={font=\sffamily\small}] 
			(S)  edge [->,double=black] node[above] {$z[k]$} (R)
			(R)  edge [->,double=black] node[right] {\hspace{.1in}$\hat{x}^{(k)}(y[0\mathbin{:}k],z[0\mathbin{:}k])$} (h)
			(x)  edge [->,double=black] (S)
			(t)  edge [->,double=black] (S)
			(I)  edge [->,double=black] node[right] {$y[k]$} (R)
			(T1) edge [- ,dashed,double=black] (T2)
			(T2) edge [->,dashed,double=black] (S);
\end{tikzpicture}
\vspace{-.1in}
\caption{\label{fig:diagram:dynamic:0} The communication structure between the strategic sensor $S$ and the side channel with the receiver $R$ for the dynamic case. Similarly, we have used a dashed edge for connecting the side-channel information to the strategic sensor to portray the fact that, for some equilibria, the sensor does not utilize its knowledge of the side-channel information and, hence, this assumption is not necessary in the framework.} 
\end{figure}
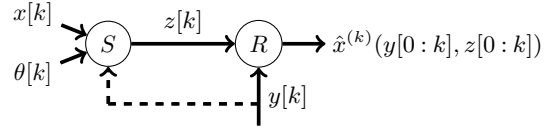

Consider an estimation problem in which the sensor and the receiver are following the communication structure in Figure~\ref{fig:diagram:dynamic:0}. The goal of the receiver is to estimate the state vector $x[k]\in\mathbb{R}^{n_x}$, which is evolving according to
$$
x[k]=A_x[k]x[k-1]+w_x[k],
$$
where $(w_x[t])_{t\in\mathbb{N}_0}$ is a sequence of i.i.d Gaussian random variables with zero mean.%\footnote{{\color{red} The case where $(w_x[t])_{t\in\mathbb{N}_0}$ are not i.i.d. can be studied as a future work. }} 
The timing of the game is as follows. At each time step $k\in\mathbb{N}_0$, first, an honest but noisy side channel reveals the measurement
$$
y[k]=C_{yx}[k]x[k]+C_{y\theta}[k]\theta[k]+w_y[k],
$$ 
where $(w_y[t])_{t\in\mathbb{N}_0}$ is a sequence of i.i.d Gaussian random variables with zero mean. Then, the strategic sensor $S$ transmits its message $z[k]$. Let $\theta[k]\in\mathbb{R}^{n_x}$ denote the private information of the sensor at time step $k\in\mathbb{N}_0$. The message that the sensor is transmitting at each time step can potentially be a function of all the previous state measurements and private signals according to the conditional distribution $p(\cdot|x[0\mathbin{:}k],y[0\mathbin{:}k],\theta[0\mathbin{:}k])$. Similar to Subsection~\ref{sec:singlesensor:static} in order to greatly simplify the presentation, we denote this by a stochastic mapping $z[k]=\gamma^{(k)}(x[0\mathbin{:}k],y[0\mathbin{:}k],\theta[0\mathbin{:}k])$ such that
\begin{align*}
\mathbb{P}\{&\gamma^{(k)}(x[0\mathbin{:}k],y[0\mathbin{:}k],\theta[0\mathbin{:}k])\in\mathcal{Z}\}
\\&=\int_{z'\in\mathcal{Z}} p(z'|x[0\mathbin{:}k],y[0\mathbin{:}k],\theta[0\mathbin{:}k]) \mathrm{d}z', \,\, \forall \mathcal{Z}\subseteq \mathbb{R}^{n_z}.
\end{align*}
Let the set of all such mappings be denoted by $\Gamma^{(k)}$. After that, the receiver calculates the best estimate of the state by \farhad{minimizing $\mathbb{E}\{\|x[k]-\farhad{\upsilon}^{(k)}(y[0\mathbin{:}k],z[0\mathbin{:}k])\|_2\}$ over $\Upsilon^{(k)}$, which} is the set of all Lebesgue-measurable functions from $\prod_{t=0}^{k}\mathbb{R}^{n_y}\times \prod_{t=0}^{k}\mathbb{R}^{n_z}$ to $ \mathbb{R}^{n_x}$. Finally, the cost functions of the receiver and the sensor for that time step are realized. We assume that the private information of the sensor is also evolving according to the linear update rule 
$$
\theta[k]=A_\theta[k]\theta[k-1]+w_\theta[k],
$$
where $(w_\theta[t])_{t\in\mathbb{N}_0}$ is a sequence of i.i.d Gaussian random variables with zero mean. We make the following standing assumption.

\begin{assumption} For each $k\in\mathbb{N}_0$, the random variables $w_x[k]$, $w_\theta[k]$, and $w_y[k]$ are jointly distributed Gaussian random variables with zero mean and a covariance matrix that satisfies
\begin{align*}
&\mathbb{E}\left\{\matrix{c}{w_x[k] \\ w_\theta[k]\\ w_y[k]}\matrix{c}{w_x[k] \\ w_\theta[k]\\ w_y[k]}^\top \right\}
\\& \hspace{.1in}=\matrix{ccc}{
\V_{w_x[k]w_x[k]} & 0 & 0 \\
0 & \V_{w_\theta[k]w_\theta[k]} & 0 \\
0 & 0 & \V_{w_y[k]w_y[k]}}\in\mathcal{S}_{++}^{2n_x+n_y}.
\end{align*}
\end{assumption}

With these definitions in hand, we are ready to define the equilibrium.

\begin{definition}\textsc{(Equilibrium):} \label{def:dynamic} A tuple $((\hat{x}^{(k)*})_{k\in \mathbb{N}_0},\linebreak(\gamma^{(k)*})_{k\in \mathbb{N}_0}) \in\prod_{k\in \mathbb{N}_0}\farhad{\mathcal{C}(\Gamma^{(k)},\Upsilon^{(k)})}\times\prod_{k\in \mathbb{N}_0}\Gamma^{(k)}$ constitutes an equilibrium for the repeated game if for all $k\in\mathbb{N}_0$, \farhad{condition~\eqref{eqn:def:dynamic}, on top of the next page, holds with}
\begin{figure*}
\begin{subequations} \label{eqn:def:dynamic}
\begin{align}
\hat{x}^{(k)*}&\in\argmin_{\hat{x}^{(k)}\in\farhad{\mathcal{C}(\Gamma^{(k)},\Upsilon^{(k)})}} \mathbb{E}\{\|x[k]-\farhad{[\hat{x}^{(k)}(\gamma^{(k)*})]}(y[0\mathbin{:}k],z[0\mathbin{:}k])\|_2\}, \\ 
\gamma^{(k)*} &\in \argmin_{\gamma^{(k)}\in\Gamma^{(k)}}\mathbb{E}\{\|(x[k]+\theta[k])  -\farhad{[\hat{x}^{(k)*}(\gamma^{(k)})]}(y[0\mathbin{:}k],z[0\mathbin{:}k-1],\gamma^{(k)}(x[0\mathbin{:}k],y[0\mathbin{:}k],\theta[0\mathbin{:}k]))\|_2\},
\end{align}
\end{subequations}
\hrule
\end{figure*}
%where 
$z[t]=\gamma^{(t)*}(x[0\mathbin{:}t],y[0\mathbin{:}t],\theta[0\mathbin{:}t])$ for all $t\in\mathbb{N}_0$.
\end{definition}

\begin{remark} Note that this definition implies that the receiver and the sensor care about their immediate cost at each time step and are, hence, myopic decision makers at each time step. This definition differs from that of a subgame perfect equilibrium (see~\cite{osborne2009introduction}) in a dynamic game in which the decision makers optimize their cost-to-go, e.g., a discounted summation of their cost over the rest of the horizon. These equilibria, referred to as myopic Nash equilibria or period-by-period Nash equilibria, have been used in the economics literature to model various competitive scenarios~\cite{ghemawat1990devolution}. Our interest in this equilibrium concept is motivated by two main factors:
\begin{itemize}
\item[(\textit{i})] In estimation theory, the optimal least mean squares filter for dynamic problems (i.e., Kalman filters) are designed so as to minimize the estimation error variance in each time step individually~\cite{kailath2000linear};
\item[(\textit{ii})] In cyber-security problems, a malicious agent might inject false data so that the state estimated by the receiver tracks that of another legitimate-looking one (while the agent is pursuing its agenda through manipulating the actual state of the system in an stealth manner)~\cite{mo2012cyber}. Hence, the agent might wish to minimize the distance between the estimate and the state of another system at each iteration separately.
\end{itemize}
\end{remark}

\begin{theorem} \label{tho:infinite} There exists at least one equilibrium in which the receiver uses the LMS estimator
\begin{align} \label{eqn:strategy:reciever}
\farhad{[\hat{x}^{(k)*}(\gamma^{(k)})]}&(y[0\mathbin{:}k],z[0\mathbin{:}k])=\matrix{cc}{\V_{x[k]\psi[k]}&\V_{x[k] z[k]}}\nonumber
\\ &\times\matrix{cc}{\V_{\psi[k]\psi[k]}&\V_{\psi[k]z[k]}\\ \V_{z[k]\psi[k]}&\V_{z[k]z[k]}}^{-1}
 \matrix{c}{\psi[k] \\ z[k]},
\end{align}
where
$$
\psi[k]=\matrix{cccccc}{y[0]^\top & \cdots & y[k]^\top & z[0]^\top & \cdots & z[k-1]^\top}^\top,
$$
and the sensor uses the policy
\begin{align}
\gamma^{(k)*}&(x[0\mathbin{:}k],y[0\mathbin{:}k],\theta[0\mathbin{:}k]) \nonumber\\
&=C_{zx}[k]x[k]+C_{z\psi}[k]\psi[k]+C_{z\theta}[k]\theta[k]+v[k].
\end{align}
In the sensor's policy, we have
\begin{align*} \label{eqn:calculatingC}
&\matrix{ccc}{C_{zx}[k] & C_{z\theta}[k] & C_{z\psi}[k]} \nonumber
\\ &\hspace{.1in}=
\matrix{c}{\V_{x[k]z[k]} \\ \V_{\theta[k]z[k]} \\ \V_{\psi[k]z[k]}}^\top
\matrix{ccc}{\V_{x[k]x[k]} &  \V_{x[k]\theta[k]} & \V_{x[k]\psi[k]}\\
\V_{\theta[k]x[k]} & \V_{\theta[k]\theta[k]} & \V_{\theta[k]\psi[k]} \\
\V_{\psi[k]x[k]} & \V_{\psi[k]\theta[k]} & \V_{\psi[k]\psi[k]} }^{-1},
\end{align*}
and $\{v[t]\}_{t\in\mathbb{N}_0}$ is a sequence of i.i.d Gaussian random variables with zero mean and covariance matrices
$$
\matrix{ccc}{\V_{v[k]x[k]}&\V_{v[k]\psi[k]}&\V_{v[k]\theta[k]}}=\matrix{ccc}{0&0&0},
$$
$$
V_{v[k]v[k]}=I-\matrix{c}{\V_{x[k]z[k]} \\ \V_{\theta[k]z[k]} \\ \V_{\psi[k]z[k]}}^\top Q[k] \matrix{c}{\V_{x[k]z[k]} \\ \V_{\theta[k]z[k]} \\ \V_{\psi[k]z[k]}},
$$
where
\begin{align*}
\matrix{c}{\V_{x[k]z[k]} \\ \V_{\theta[k]z[k]} \\ \V_{\psi[k]z[k]}}\in
\argmin_{\xi\in\mathbb{R}^{(2n_x+\dim(\psi[k]))\times n_z}} & \trace ( \xi^\top W[k] \xi) \\
\mathrm{s.t.} \hspace{.47in} & \,\,\xi^\top Q[k] \xi \leq I,
\end{align*}
with 
\begin{equation*} 
\begin{split}
W[k]&=\Xi[k]^\top\matrix{cc}{I & -I \\ -I & 0}\Xi[k],
\end{split}
\end{equation*}
\begin{equation*} \label{eqn:definition:Q}
\begin{split}
Q[k]&=\Xi[k]^\top \bigg(\matrix{ccc}{\V_{x[k]x[k]} & \V_{x[k]\theta[k]} \\ \V_{\theta[k]x[k]} & \V_{\theta[k]\theta[k]}}\\&-\matrix{c}{\V_{x[k]\psi[k]} \\ \V_{\theta[k]\psi[k]}}\V_{\psi[k]\psi[k]}^{-1} \matrix{c}{\V_{x[k]\psi[k]} \\ \V_{\theta[k]\psi[k]}}^\top\bigg)^{-1}\Xi[k],
\end{split}
\end{equation*}
$$
\Xi[k]=\matrix{ccc}{I & 0 & -\V_{x[k]\psi[k]}\V_{\psi[k]\psi[k]}^{-1}\\ 0 & I & -\V_{\theta[k]\psi[k]}\V_{\psi[k]\psi[k]}^{-1}}.
$$
Furthermore, the sensor's policy of the form $(\kappa_k\gamma^{(k)*})_{k\in\mathbb{N}_0}$, for $\kappa_k\in \mathbb{R}\setminus\{0\},\forall k\in\mathbb{N}_0$, along side the receiver's policy $(\hat{x}^{(k)*})_{k\in\mathbb{N}_0}$, also constitutes an equilibrium. All these equilibria result in the same estimation error variance at the receiver.
\end{theorem}

\begin{IEEEproof} The proof follows from applying the results of Theorem~\ref{tho:1} in each time step and treating all the accumulated information at this time $y[0],\dots,y[k],z[0],\dots,z[k-1]$ as the side-channel information. \end{IEEEproof}

The strategies of the receiver and the sensor in the portrayed equilibria in Theorem~\ref{tho:infinite} can potentially require an infinite amount of memory because the size $\psi[k]$ grows with $k$. However, similar to the previous subsection, we would like to find an equilibrium for which $C_{z\psi}[k]=0$ and, hence, the sensor does require an infinite memory to keep track of all the previously transmitted signals. This is discussed in the following corollary.

\begin{corollary} \label{cor:memoryless} There exists at least one equilibrium in which $\gamma^{(k)*}$ is a memory-less function for all $k\in\mathbb{N}_0$. More precisely, the receiver uses the LMS estimator in
\begin{equation} \label{eqn:strategy:reciever:1}
\farhad{[\hat{x}^{(k)*}(\gamma^{(k)})]}(y[0\mathbin{:}k],z[0\mathbin{:}k])=\mathbb{E}\{x[k]\,|\,y[0\mathbin{:}k],z[0\mathbin{:}k]\},
\end{equation}
and the sensor uses the policy
\begin{align*}
\gamma^{(k)*}(x[0\mathbin{:}k],&y[0\mathbin{:}k],\theta[0\mathbin{:}k])\\
&=C_{zx}[k]x[k]+C_{z\theta}[k]\theta[k]+v[k],
\end{align*}
where $\{v[t]\}_{t\in\mathbb{N}_0}$ is a sequence of i.i.d Gaussian random variables with zero mean and covariance matrices
$$
\matrix{ccc}{\V_{v[k]x[k]}&\V_{v[k]\psi[k]}&\V_{v[k]\theta[k]}}=\matrix{ccc}{0&0&0},
$$
$$
V_{v[k]v[k]}=I-\matrix{cc}{C_{zx}[k] & C_{z\theta}[k]} \Xi'[k] \matrix{c}{C_{zx}[k]^\top \\ C_{z\theta}[k]^\top},
$$
in which 
\begin{align*}
\matrix{c}{C_{zx}[k]^\top \\ C_{z\theta}[k]^\top}\in
\argmin_{\xi'\in\mathbb{R}^{2n_x\times n_z}} & \trace ( \xi^{'\top} W'[k] \xi') \\
\mathrm{s.t.} \hspace{.2in} & \,\,\xi^{'\top} \Xi'[k] \xi' \leq I,
\end{align*}
with 
\begin{align*}
W'[k]=&\Xi'[k]^\top\matrix{cc}{-I & -I \\ -I & 0}\Xi'[k], \\
\Xi'[k]=&\matrix{ccc}{\V_{x[k]x[k]} & \V_{x[k]\theta[k]} \\ \V_{\theta[k]x[k]} & \V_{\theta[k]\theta[k]}}
\\&-\matrix{c}{\V_{x[k]\psi[k]} \\ \V_{\theta[k]\psi[k]}}\V_{\psi[k]\psi[k]}^{-1} \matrix{c}{\V_{x[k]\psi[k]} \\ \V_{\theta[k]\psi[k]}}^\top.
\end{align*}
Furthermore, the sensor's policy of the form $(\kappa_k\gamma^{(k)*})_{k\in\mathbb{N}_0}$, for $\kappa_k\in \mathbb{R}\setminus\{0\},\forall k\in\mathbb{N}_0$, along side the receiver's policy $(\hat{x}^{(k)*})_{k\in\mathbb{N}_0}$, also constitutes an equilibrium. All these equilibria result in the same estimation error variance at the receiver.
\end{corollary}

\begin{IEEEproof} The proof follows from applying the results of Corollary~\ref{cor:notafunctionofy} at each time step.
\end{IEEEproof}

Now that we have showed that there is at least one equilibrium in which the measurement $z[k]$ is constructed using an affine memory-less mapping, we can generate a recursive filter for constructing the best estimate in~\eqref{eqn:strategy:reciever:1}. The following remark is devoted to this construction.

\begin{remark} \label{remark:Kalman} The receiver needs to implement a Kalman filter (e.g., see~\cite{kailath2000linear}) for the equilibrium in Corollary~\ref{cor:memoryless}. To do so, first, with slight abuse of notation, let us introduce 
$$
\matrix{c}{\hat{x}[k] \\ \hat{\theta}[k]}=\matrix{c}{\mathbb{E}\{x[k]|y[0\mathbin{:}k],z[0\mathbin{:}k]\} \\ \mathbb{E}\{\theta[k]|y[0\mathbin{:}k],z[0\mathbin{:}k]\}}, \forall k\in\mathbb{N}_0.
$$
Hence, clearly, $\hat{x}^{(k)*}(y[0\mathbin{:}k],z[0\mathbin{:}k])=\hat{x}[k]$. Now, we may also define the error covariance matrix
$$
P[k]=\mathbb{E}\left\{\matrix{c}{x[k]-\hat{x}[k] \\ \theta[k]-\hat{\theta}[k]} \matrix{c}{x[k]-\hat{x}[k] \\ \theta[k]-\hat{\theta}[k]}^\top \right\}, \forall k\in\mathbb{N}_0.
$$
Transitioning from time step $k-1$ to time step $k$, the first stage of the Kalman filter is the prediction phase, which results in the estimate
\begin{align*}
\matrix{c}{\tilde{x}[k] \\ \tilde{\theta}[k]}
&=\matrix{c}{\mathbb{E}\{x[k]|y[0\mathbin{:}k-1],z[0\mathbin{:}k-1]\} \\ \mathbb{E}\{\theta[k]|y[0\mathbin{:}k-1],z[0\mathbin{:}k-1]\}}\\
&=\matrix{cc}{A_x[k] & 0 \\ 0 & A_\theta[k]}\matrix{c}{\hat{x}[k-1] \\ \hat{\theta}[k-1]},
\end{align*}
and the error covariance update
\begin{align*}
\tilde{P}[k]
&=\mathbb{E}\left\{\matrix{c}{x[k]-\tilde{x}[k] \\ \theta[k]-\tilde{\theta}[k]} \matrix{c}{x[k]-\tilde{x}[k] \\ \theta[k]-\tilde{\theta}[k]}^\top \right\}\\
&=\matrix{cc}{A_x[k] & 0 \\ 0 & A_\theta[k]}P[k-1]\matrix{cc}{A_x[k] & 0 \\ 0 & A_\theta[k]}^\top
\\&\hspace{.1in}+\matrix{cc}{\V_{w_x[k]w_x[k]} & 0 \\ 0 & \V_{w_\theta[k]w_\theta[k]}}.
\end{align*}
After receiving the measurement $y[k]$, i.e., the information shared by the side channel, we may update the estimate to
\begin{align*}
\matrix{c}{\grave{x}[k] \\ \grave{\theta}[k]}
&=\matrix{c}{\mathbb{E}\{x[k]|y[0\mathbin{:}k],z[0\mathbin{:}k-1]\} \\ \mathbb{E}\{\theta[k]|y[0\mathbin{:}k],z[0\mathbin{:}k-1]\}}\\
&=\matrix{c}{\tilde{x}[k] \\ \tilde{\theta}[k]}+\grave{K}[k]\bigg(y[k]-C_y[k]\matrix{c}{\tilde{x}[k] \\ \tilde{\theta}[k]}\bigg),
\end{align*}
where 
$$
\grave{K}[k]=\tilde{P}[k]C_y[k]^\top(C_y[k]\tilde{P}[k]C_y[k]^\top+\V_{w_y[k]w_y[k]})^{-1},
$$
with $C_y[k]=\matrix{cc}{C_{yx}[k] & C_{y\theta}[k]}$. This update improves the error covariance matrix according to
\begin{align*}
\grave{P}[k]
&=\mathbb{E}\left\{\matrix{c}{x[k]-\grave{x}[k] \\ \theta[k]-\grave{\theta}[k]} \matrix{c}{x[k]-\grave{x}[k] \\ \theta[k]-\grave{\theta}[k]}^\top \right\}\\
&=(I-\grave{K}[k]C_y[k])\tilde{P}[k].
\end{align*}
Finally, after receiving the measurement $z[k]$, i.e., the information transmitted by the strategic sensor, we may update the estimate to
\begin{align*}
\matrix{c}{\hat{x}[k] \\ \hat{\theta}[k]}
&=\matrix{c}{\grave{x}[k] \\ \grave{\theta}[k]}+K[k]\left(z[k]-C_z^*[k]\matrix{c}{\grave{x}[k] \\ \grave{\theta}[k]}\right),
\end{align*}
where
\begin{align*}
K[k]&=\grave{P}[k]C_z[k]^\top(C_z[k]\grave{P}[k]C_z[k]^\top+\V_{w_z[k]w_z[k]})^{-1},
\end{align*}
with $C_z[k]=\matrix{cc}{C_{zx}[k] & C_{z\theta}[k]}$.  This update results in the error covariance update rule
\begin{align*}
P[k]
&=(I-K[k]C_z[k])\grave{P}[k].
\end{align*}
Therefore, we have a recursive scheme for constructing the estimates of the receiver. Moreover, calculating $\Xi'[k]$ is also straightforward using the parameters of the introduced Kalman filter as
\begin{align*}
\Xi'[k]=&\matrix{ccc}{\V_{x[k]x[k]} & \V_{x[k]\theta[k]} \\ \V_{\theta[k]x[k]} & \V_{\theta[k]\theta[k]}}
\\&-\matrix{c}{\V_{x[k]\psi[k]} \\ \V_{\theta[k]\psi[k]}}\V_{\psi[k]\psi[k]}^{-1} \matrix{c}{\V_{x[k]\psi[k]} \\ \V_{\theta[k]\psi[k]}}^\top
\\
%=&\mathbb{E}\left\{\hspace{-.03in}\matrix{c}{\hspace{-.05in} x[k]-\mathbb{E}\{x[k]|\psi[k]\}\hspace{-.05in} \\ \hspace{-.05in}\theta[k]-\mathbb{E}\{\theta[k]|\psi[k]\}\hspace{-.05in}} \matrix{c}{\hspace{-.05in} x[k]-\mathbb{E}\{x[k]|\psi[k]\}\hspace{-.05in} \\ \hspace{-.05in}\theta[k]-\mathbb{E}\{\theta[k]|\psi[k]\}\hspace{-.05in}}^{\hspace{-.03in}\top} \hspace{-.03in}\right\}\\
=&\mathbb{E}\left\{\matrix{c}{x[k]-\grave{x}[k] \\ \theta[k]-\grave{\theta}[k]} \matrix{c}{x[k]-\grave{x}[k] \\ \theta[k]-\grave{\theta}[k]}^\top \right\}=\grave{P}[k].
\end{align*}
\end{remark}

\section{Multiple sensors} \label{sec:multiplesensor}
We now consider static estimation with multiple strategic sensors for both synchronous and asynchronous communication.

\subsection{Static Estimation with Synchronous Independent Sensors}

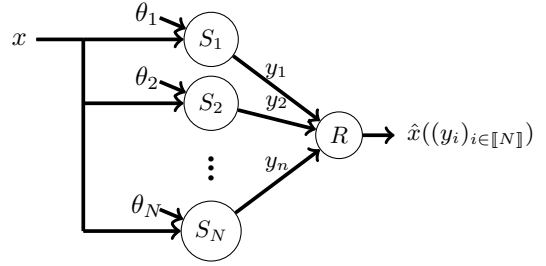
\begin{figure}[t]
\centering
\begin{tikzpicture}[scale=0.85]
\node[state,minimum size=0.1cm,scale=0.9] (R)  at (+1.0,+0.0) {$R$};
\node[state,minimum size=0.1cm,scale=0.9] (S1) at (-1.0,+1.5) {$S_1$};
\node[state,minimum size=0.1cm,scale=0.9] (S2) at (-1.0,+0.5) {$S_2$};
\node[minimum size=0.1cm,scale=0.9] 	  (S3) at (-1.0,-0.4) {\huge $\vdots$};
\node[state,minimum size=0.1cm,scale=0.9] (Sn) at (-1.0,-1.5) {$S_N$};
\node (xr) at (-4.0,+1.5) {$x$};
\node[coordinate] (x) at (-3.0,+1.5) {};
\node[coordinate] (xp2) at (-3.0,-1.5) {};
\node[coordinate] (xp1) at (-3.0,+0.5) {};
\node (t1p) at (-2.0,+1.9) {$\theta_1$};
\node (t2p) at (-2.0,+0.9) {$\theta_2$};
\node (t3p) at (-2.0,-1.1) {$\theta_N$};
\node (t1) at (-1.95,+1.9) {};
\node (t2) at (-1.95,+0.9) {};
\node (t3) at (-1.95,-1.1) {};
\node[minimum size=0.01cm,scale=0.9] (h)  at (+2.0,-0.0 ) {};
\path[every node/.style={font=\sffamily\small}] 
		(R)  edge [->,double=black] node[right] {\hspace{.1in}$\hat{x}((y_i)_{i\in\N})$} (h)
		(t1)   edge [->,double=black] node[below] {}          (S1)
		(t2)   edge [->,double=black] node[below] {}          (S2)
		(t3)   edge [->,double=black] node[below] {}          (Sn)
		(S1)  edge [->,double=black] node[above] {$y_1$} (R)
		(S2)  edge [->,double=black] node[above] {$y_2$} (R)
		(x)    edge [->,double=black] node[below] {}          (S1)
		(xp2)edge [->,double=black] node[below] {}          (Sn)
		(xp1)edge [->,double=black] node[below] {}          (S2)
		(x)    edge [-,  double=black] node[below] {}          (xp2)				
		(xr)   edge [-,  double=black] node[below] {}          (x)				
		(Sn)  edge [->,double=black] node[above] {$y_n$} (R);
\end{tikzpicture}
\vspace{-.1in}
\caption{\label{fig:1} The communication structure between the sensors $(S_i)_{i\in\N}$ and the receiver $R$.}
\end{figure}

In this section, we assume that the sensors and the receiver are connected to each other using the communication network presented in Figure~\ref{fig:1}. For each $i\in \N$, sensor $S_i$ transmits the message $y_i\in\mathbb{R}^{n_{y_i}}$ to the receiver $R$. All these signals are transmitted at exactly the same time  through parallel secure communication channels to the receiver (see Figure~\ref{fig:1}). Hence, the sensors do not have access to each other messages and cannot use this information for constructing their signals. Note that the parameters of the policies of other sensors may be available but the signal realization itself is off limit. Again, as in the last section, each sensor $S_i$ has access to the exact value of the state $x\in\mathbb{R}^{n_x}$ (which the receiver wants to estimate) and its private parameter $\theta_i\in\mathbb{R}^{n_x}$. We also assume that $x$ and $\theta_i$, $i\in\N$, are jointly distributed Gaussian random variables with zero mean. 

Motivated by the results of the previous section, which showed that there exists an equilibrium in which the sensor uses an affine policy, we now \textit{restrict} ourselves to affine sensor policies in the multi-sensor case. More precisely, we assume that sensor $i$'s policy $\gamma_i$ is of the form
\begin{align*}
y_i&=\gamma_i(x,\theta_i)\\&=a_i^\top x+b_i^\top \theta_i+v_i,
\end{align*}
where $a_i,b_i\in\mathbb{R}^{n_x\times n_{y_i}}$ are deterministic values and $v_i\in\mathbb{R}^{n_{y_i}}$ is a zero mean Gaussian random variable. The set  of all such policies is denoted by $\Gamma_i$.  In these policies, without loss of generality, we can assume that for any $i\in\N$, $v_i$ is statistically independent of $x$ and $\theta_i$. Note that two \textit{jointly} distributed Gaussian random variables are statistically independent if and only if they are uncorrelated~\cite[p.108]{1979stochastic}. To show that this assumption is without loss of generality, suppose that $v_i$ is not statistically independent of $x$ and/or $\theta_i$, that is, $\V_{v_ix}\neq 0$ and/or $\V_{v_i\theta_i}\neq 0$. In such case, we can clearly rewrite the output vector as
$$
y_i={a'_i}^\top x+{b'_i}^\top \theta_i+v'_i,
$$
where 
$$
a'_i=a_i+\left(\matrix{c}{\V_{v_ix}\\ \V_{v_i\theta_i}}\matrix{cc}{\V_{xx} & \V_{x\theta_i} \\ \V_{\theta_ix} & \V_{\theta_i\theta_i}}^{-1}\matrix{c}{I\\ 0}\right)^{\hspace{-.07in}\top}\hspace{-.05in},
$$
$$
b'_i=b_i+\left(\matrix{c}{\V_{v_ix}\\ \V_{v_i\theta_i}}\matrix{cc}{\V_{xx} & \V_{x\theta_i} \\ \V_{\theta_ix} & \V_{\theta_i\theta_i}}^{-1}\matrix{c}{0\\ I}\right)^{\hspace{-.07in}\top}\hspace{-.05in},
$$
and
$$
v'_i=v_i-\matrix{c}{\V_{v_ix}\\ \V_{v_i\theta_i}}\matrix{cc}{\V_{xx} & \V_{x\theta_i} \\ \V_{\theta_ix} & \V_{\theta_i\theta_i}}^{-1}\matrix{c}{x\\ \theta_i} \hspace{-.05in}.
$$
For this new representation, it follows from simple algebraic manipulations that $\V_{v'_ix}=0$ and $\V_{v'_i\theta_i}=0$, i.e., that $v'_i$ is independent of $x$ and $\theta_i$ (again, since all these variables are jointly distributed Gaussian random variables).

Each $\gamma_i\in\Gamma_i$ can be equivalently represented using the tuple $(a_i,b_i,\V_{v_iv_i})$ and, therefore, the set of feasible policies $\Gamma_i$ is isomorphic to the product space $\mathbb{R}^{n_x\times n_{y_i}}\times \mathbb{R}^{n_x\times n_{y_i}} \times \mathcal{S}_{+}^{n_{y_i}}$. Taking advantage of this bijection, we will sometimes abuse notation and refer to this tuple directly as $\gamma_i$.

Following the transmission of messages $y=(y_i)_{i\in\N}$, the receiver computes $\farhad{\upsilon(y)=}\mathbb{E}\{x\,|\,y_1,\dots,y_N\}$ so as to minimize $\mathbb{E}\{\|x-x'\|_2^2\}$ over the set of random variables $x'$ measurable with respect to $y_1,\dots,y_N$.

Similar to the problem formulation of the last section, the ultimate goal of each sensor $S_i$, $i\in\N$, is to make sure $\farhad{\upsilon}(y)$ is a good estimate of $x$ \emph{plus its private information $\theta_i$}. Therefore, the cost function that sensor~$i$, $i\in\N$, is trying to minimize is 
$$
\mathbb{E}\{\|(x+\theta_i)-\farhad{\upsilon}(\gamma_i(x,\theta_i),(\gamma_j(x,\theta_j))_{j\neq i})\|_2^2\}.
$$
This naturally leads us to the following definition.

\begin{definition}\textsc{(Equilibrium in Affine Strategies):} \label{def:generalequi} Let $\Upsilon$ denote the set of all Lebesgue-measurable functions from $\prod_{i\in\N}\mathbb{R}^{n_{y_i}}$ to $\mathbb{R}^{n_x}$. A tuple $(\hat{x}^*\hspace{-.03in},\hspace{-.03in}(\gamma_i^*)_{i\in\N}\hspace{-.03in}) \linebreak\in\farhad{\mathcal{C}(\prod_{i\in\N}\Gamma_i,\Upsilon)}\times\prod_{i\in\N}\Gamma_i$ constitutes an equilibrium in affine strategies if 
\begin{subequations}
\begin{align}
\hat{x}^* &\in\hspace{-.23in}\argmin_{\hat{x}\in\farhad{\mathcal{C}(\prod_{i\in\N}\Gamma_i,\Upsilon)}}\hspace{-.23in} \mathbb{E}\{\|x-\farhad{[\hat{x}((\gamma^*_j)_{j\in\N})]}((\gamma_j^*(x,\theta_j))_{j\in\N})\|_2^2\}, \\ 
\gamma_i^* &\in \argmin_{\gamma_i\in\Gamma_i} \mathbb{E}\{\|(x+\theta_i)-\farhad{[\hat{x}^*(\gamma_i,(\gamma_j^*)_{j\neq i})]}\nonumber\\&\hspace{1.2in}(\gamma_i(x,\theta_i),(\gamma^*_j(x,\theta_j))_{j\neq i})\|_2^2\},
\end{align}
\end{subequations}
for all $i\in\N$.
\end{definition}

\begin{remark} \label{remark:linear_nonlinear}
Note that the qualifier ``in affine strategies'' in the definition above means that the equilibrium in question is a best response \emph{only} when the sensors' strategy space is the set of all affine policies $\Gamma_i$. This analysis gives a lower-bound on the influence of the sensors (on the quality of the estimation) since they find more degrees of freedom for constructing untruthful messages (and, hence, create a larger error to their benefit) by extending their set of strategies to also cover nonlinear mappings.
\end{remark}

We are interested in situations where the sensors population is large and homogeneous, at least as perceived by the receiver. In this case, it is natural to model the private parameters $\theta_i$, $i\in\N$, as i.i.d random variables. Keeping in mind that we assumed these are jointly distributed Gaussian random variables with zero mean, their distribution is fully characterized by their covariance matrices $\V_{x\theta_i}=\V_{x\theta}$ for all $i\in\N$. Furthermore, let $\V_{\theta_i\theta_j}=\V_{\theta\theta}$ if $i=j$ and $\V_{\theta_i\theta_j}=\U_{\theta\theta}$ otherwise. In this homogeneous context, the receiver should expect all sensors to use the same policy, and the most compelling characterization of the population's behavior is thus provided by symmetric equilibria.
In the remainder of this subsection, we show that such a symmetic equilibrium in affine strategies indeed exists. To do so, we first need to prove the following lemma.

\begin{lemma} \label{lemma:1} If $\gamma_i=(a,b,\V_{vv})$ for all $i\in\N$, then
$\mathbb{E}\{x|y_1,\dots,y_N\}=\mathbb{E}\{x|(y_1+\dots+y_N)/N\}.$
\end{lemma}

\begin{IEEEproof} See Appendix~\ref{proof:lemma:1}.
\end{IEEEproof}

We are now in a position to prove the main result of this part regarding the existence of symmetric equilibria in affine strategies. In order to derive explicit expressions, we henceforth assume that $\dim(y_i)=1$ for all $i\in\N$, i.e., that all sensors use scalar messages. 

\begin{theorem} \label{tho:general} Assume that $n_{y_i}=1$ for all $i\in\N$, $\V_{x\theta}=0$, and $\U_{\theta\theta}=0$. There exists a symmetric equilibrium in affine strategies in which the receiver follows $\farhad{[\hat{x}^*((\gamma_i)_{i\in\N})]}(y)=\mathbb{E}\{x|(y_1+\dots+y_N)/N\}$ and sensor $S_i$, $i\in\N$, employs the linear policy $\gamma^*= (a^*,b^*,0)$ where
\begin{equation*}
\begin{split}
\matrix{c}{b^*\\ a^*}=\sqrt{\frac{1}{1+(N-1)\xi_1^\top \xi_1}}\matrix{cc}{N\V_{\theta\theta}^{-1/2} & 0 \\ 0 & \V_{xx}^{-1/2}}\xi
\end{split}
\end{equation*}
and $\xi=\matrix{cc}{\xi_1^\top & \xi_2^\top}^\top$ is the normalized eigenvector (i.e., $\|\xi\|_2=1$) corresponding to the smallest eigenvalue of the matrix
\begin{align*}
\matrix{cc}{0 & -\V_{\theta\theta}^{1/2}\V_{xx}^{1/2} \\ -\V_{xx}^{1/2}\V_{\theta\theta}^{1/2} & -\V_{xx}}.
\end{align*}
Furthermore, the sensors' policy of the form $\kappa\gamma^*$, for some $\kappa\in \mathbb{R}\setminus\{0\}$, along side the receiver's policy $\hat{x}^*$, also constitutes a symmetric equilibrium in affine strategies.
\end{theorem}

\begin{IEEEproof} Let $\bar{y}=N^{-1}\sum_{i=1}^N y_i$, $\bar{a}=N^{-1}\sum_{i=1}^N a_i$, and $\bar{v}=N^{-1}\sum_{i=1}^N v_i$. \farhad{For the sake of the simplicity of the presentation, in this proof, we write $\hat{x}^*(y)$ instead of $[\hat{x}^*((\gamma_i)_{i\in\N})](y)$.} For calculating $\hat{x}\farhad{^*}(y)=\V_{x\bar{y}}\V_{\bar{y}\bar{y}}^{-1}\bar{y}$, first, we need to compute the following quantities
\begin{align} 
\V_{x\bar{y}}&=\mathbb{E}\left\{x\bar{y}^\top\right\}\nonumber\\
&=\mathbb{E}\left\{x\left[ \bar{a}^\top x +\frac{1}{N}\sum_{i=1}^N b_i^\top \theta_i+\bar{v} \right]^{\hspace{-.05in}\top}\right\}\nonumber\\
&=\V_{xx} \bar{a}, \label{eqn:expansion:Vxbary}
\end{align}
and
\begin{align}
\V_{\bar{y}\bar{y}}&\hspace{-.03in}=\hspace{-.03in}\mathbb{E}\left\{\bar{y}\bar{y}^\top\right\} \nonumber \\
&\hspace{-.03in}=\hspace{-.03in}\mathbb{E}\Bigg\{\hspace{-.05in}\left[ \bar{a}^\top \hspace{-.03in}x \hspace{-.03in}+\hspace{-.03in}\frac{1}{N}\sum_{i=1}^N b_i^\top \theta_i\hspace{-.03in}+\hspace{-.03in}\bar{v} \right]\hspace{-.07in}\left[ \bar{a}^\top \hspace{-.03in}x \hspace{-.03in}+\hspace{-.03in}\frac{1}{N}\sum_{i=1}^N b_i^\top \theta_i\hspace{-.03in}+\hspace{-.03in}\bar{v} \right]^{\hspace{-.05in}\top}\hspace{-.07in}\Bigg\} \nonumber \\
&\hspace{-.03in}=\hspace{-.03in}\bar{a}^\top\V_{xx} \bar{a}+\frac{1}{N^2}\sum_{i=1}^N b_i^\top \V_{\theta\theta} b_i+\frac{1}{N^2}\sum_{i=1}^N \V_{v_iv_i}. \label{eqn:Vbarybary}
\end{align}
We can also expand the cost of each agent as
\begin{align} 
\mathbb{E}&\{\|(x+\theta_i)-\hat{x}\farhad{^*}(y)\|_2^2\}\nonumber\\
%\\&=
%\trace(\mathbb{E}\{((x+\theta_i)-\hat{x}\farhad{^*}(y))((x+\theta_i)-\hat{x}\farhad{^*}(y))^\top\})\nonumber\\
&=\trace(\V_{xx}+\V_{\theta\theta}-\mathbb{E}\{(x+\theta_i)\hat{x}\farhad{^*}(y)^\top\}\nonumber\\
&\;\;\;-\mathbb{E}\{\hat{x}\farhad{^*}(y)(x+\theta_i)^\top\}+\mathbb{E}\{\hat{x}\farhad{^*}(y)\hat{x}\farhad{^*}(y)^\top\}).\label{eqn:expansionofagentcost}
\end{align}
Now, notice that the identity in~\eqref{eqn:long:x+thetaandhatx} holds. 
\begin{figure*}
\begin{equation} \label{eqn:long:x+thetaandhatx}
\begin{split}
\mathbb{E}\{(x+\theta_i)\hat{x}\farhad{^*}(y)^\top\}&=\mathbb{E}\Bigg\{\hspace{-.05in}(x+\theta_i)\Bigg[ \Big(\frac{1}{N}\sum_{i=1}^N a_i\Big)^\top x +\frac{1}{N}\sum_{i=1}^N b_i^\top \theta_i+\frac{1}{N}\sum_{i=1}^N v_i \Bigg]^\top\hspace{-.05in}\V_{\bar{y}\bar{y}}^{-1}\V_{\bar{y}x}\hspace{-.05in}\Bigg\}=\left(\hspace{-.05in}\V_{x\bar{y}}+\frac{1}{N}\V_{\theta\theta}b_i\hspace{-.05in}\right)\V_{\bar{y}\bar{y}}^{-1}\V_{\bar{y}x}.
\end{split}
\end{equation}
\hrulefill
\end{figure*}
Substituting~\eqref{eqn:long:x+thetaandhatx} into~\eqref{eqn:expansionofagentcost} results in
\begin{align} 
\mathbb{E}&\{\|(x+\theta_i)-\hat{x}\farhad{^*}(y)\|_2^2\}\nonumber\\
%&=\trace\big(\V_{xx}+\V_{\theta\theta}-(\V_{x\bar{y}}+N^{-1}\V_{\theta\theta}b_i)\V_{\bar{y}\bar{y}}^{-1}\V_{\bar{y}x}\nonumber\\
%&\hspace{.1in}-\V_{x\bar{y}}\V_{\bar{y}\bar{y}}^{-1}(\V_{x\bar{y}}+N^{-1}\V_{\theta\theta}b_i)^\top+\V_{x\bar{y}}\V_{\bar{y}\bar{y}}^{-1}\V_{\bar{y}x}\big)\nonumber\\
&=\trace(\V_{xx}\hspace{-.03in}+\hspace{-.03in}\V_{\theta\theta})\hspace{-.03in}+\hspace{-.03in}\trace\big(\hspace{-.05in}-\hspace{-.05in}N^{-1}\V_{\theta\theta}b_i\V_{\bar{y}\bar{y}}^{-1}\V_{\bar{y}x}\hspace{-.03in}\nonumber\\
&\hspace{.1in}-\V_{x\bar{y}}\V_{\bar{y}\bar{y}}^{-1}N^{-1}b_i^\top\V_{\theta\theta}^\top-\V_{x\bar{y}}\V_{\bar{y}\bar{y}}^{-1}\V_{\bar{y}x}\big).\label{eqn:expansion:cost}
\end{align}
The receiver, without incurring any information loss, can scale up or down the received measurements to make sure $\V_{\bar{y}\bar{y}}=1$ (note that $\dim(\bar{y})=1$). Let us show that this assumption is without loss of generality. To do so, notice that by fixing strategies of sensors $j\neq i$, we can calculate the best response of sensor $i$ through solving 
\begin{equation*}
\min_{a_i,b_i,V_{v_iv_i}}  \mathbb{E}\{\|(x+\theta_i)-V_{x\bar{y}}V_{\bar{y}\bar{y}}^{-1}\bar{y}\|_2^2\}, 
\end{equation*}
because $\hat{x}^*(y)=(V_{x\bar{y}}/V_{\bar{y}\bar{y}})\bar{y}$ (recall that $\dim(\bar{y})=1$). Notice that $\bar{y}$ is implicitly a function of $(a_i,b_i,V_{v_iv_i})$. Now, let us define $\tilde{y}=\bar{y}/\sqrt{V_{\bar{y}\bar{y}}}$. Evidently, $V_{\tilde{y}\tilde{y}}=1$ and $V_{x\tilde{y}}=V_{x\bar{y}}/\sqrt{V_{\bar{y}\bar{y}}}$. Therefore, $\hat{x}^*(y)=V_{x\tilde{y}}\tilde{y}$ (following simple algebraic manipulations). This indeed shows that we can calculate the best response of sensor $i$ by solving
\begin{align*}
\min_{\tilde{a}_i,\tilde{b}_i,\tilde{V}_{v_iv_i}} & \mathbb{E}\{\|(x+\theta_i)-V_{x\tilde{y}}\tilde{y}\|_2^2\}, \\[-.3em]
\mathrm{s.t.} \hspace{.15in} & V_{\tilde{y}\tilde{y}}=1,
\end{align*}
in which $\tilde{a}_i=a_i/\sqrt{V_{\bar{y}\bar{y}}}$, $\tilde{b}_i=b_i/\sqrt{V_{\bar{y}\bar{y}}}$, and $\tilde{V}_{v_iv_i}=V_{v_iv_i}/V_{\bar{y}\bar{y}}$. Therefore, these two optimization problems are equivalent and, hence, the assumption that $V_{\bar{y}\bar{y}}=1$ is without loss of generality. Now, by substituting~\eqref{eqn:expansion:Vxbary} into~\eqref{eqn:expansion:cost}, we get
\begin{align*}
\mathbb{E}\{\|(x+&\theta_i)-\hat{x}\farhad{^*}(y)\|_2^2\}\\&=\trace(\V_{xx}+\V_{\theta\theta})+\trace(-N^{-1}\V_{\theta\theta}b_i\V_{\bar{y}x}\\&\hspace{.5in}-\V_{x\bar{y}}N^{-1}b_i^\top\V_{\theta\theta}^\top-\V_{x\bar{y}}\V_{\bar{y}x})\\
&=\trace(\V_{xx}+\V_{\theta\theta})+\trace\big(-N^{-1}\V_{\theta\theta}b_i\bar{a}^\top\V_{xx}\\ &\hspace{.5in}-\V_{xx} \bar{a}N^{-1}b_i^\top\V_{\theta\theta}^\top-\V_{xx} \bar{a} \bar{a}^\top\V_{xx}\big)\\&=\trace(\V_{xx}+\V_{\theta\theta})+\trace\big(-N^{-1}\bar{a}^\top\V_{xx}\V_{\theta\theta}b_i\\ &\hspace{.5in}-N^{-1}b_i^\top\V_{\theta\theta}\V_{xx} \bar{a}- \bar{a}^\top\V_{xx}\V_{xx} \bar{a}\big)\\
&=\trace(\V_{xx}+\V_{\theta\theta})+\matrix{c}{b_i \\ \bar{a}}^\top G \matrix{c}{b_i \\ \bar{a}},
\end{align*}
where 
$$
G=\matrix{cc}{0 & -\frac{1}{N}\V_{\theta\theta}\V_{xx}\\ -\frac{1}{N}\V_{xx}\V_{\theta\theta} & -\V_{xx}\V_{xx}}.
$$
Let $\beta_i:\prod_{j\neq i}\Gamma_j\rightarrow \Gamma_i$ denote the best response of player $i$. This mapping is defined as $\beta_i(\gamma_{-i})=(a_i^*,b_i^*,V_{v_iv_i}^*)$ where 
\begin{equation*}
\begin{split}
V_{v_iv_i}^*=1&-\frac{1}{N^2}\sum_{j\neq i}V_{v_jv_j}-\bar{a}^{*\top}\V_{xx} \bar{a}^*-\frac{1}{N^2}\sum_{j\neq i} b_j^\top \V_{\theta\theta} b_j\\ &-\frac{1}{N^2} b_i^{*\top} \V_{\theta\theta} b_i^*,
\end{split}
\end{equation*}
and $a_i^*=N\bar{a}^*-\sum_{j\neq i} a_j$ with
\begin{subequations} \label{eqn:appendix:optimization:1}
\begin{align}
(\bar{a}^*,b_i^*)\in\argmin_{\bar{a},b_i} & \hspace{.1in} \matrix{c}{b_i \\ \bar{a}}^\top G\matrix{c}{b_i \\ \bar{a}},\\
\mathrm{s.t.}\hspace{.1in} & \hspace{.1in} \bar{a}^\top\V_{xx} \bar{a}+\frac{1}{N^2}\sum_{j=1}^N b_j^\top \V_{\theta\theta} b_j \nonumber\\ &\hspace{.6in}\leq 1-\frac{1}{N^2}\sum_{j\neq i}V_{v_jv_j},
\label{eqn:appendix:optimization:1:constraints}
\end{align}
\end{subequations}
where the constraint~\eqref{eqn:appendix:optimization:1:constraints} is motivated by 
\begin{align*}
\V_{v_iv_i}
&=1-\bar{a}^\top\V_{xx} \bar{a}-\frac{1}{N^2}\sum_{j=1}^N b_j^\top \V_{\theta\theta} b_j-\frac{1}{N^2}\sum_{j\neq i} \V_{v_jv_j}\geq 0,
\end{align*}
which can be extracted through rearranging the terms in~\eqref{eqn:Vbarybary} while setting $\V_{\bar{y}\bar{y}}=1$. Note that $\V_{v_iv_i}\geq 0$ because the variance of any random variable, by definition, must be non-negative. Now, we can rewrite~\eqref{eqn:appendix:optimization:1} as
\begin{subequations} \label{eqn:appendix:optimization:2}
\begin{align}
\min_{\bar{a},b_i} & \hspace{.1in} \matrix{c}{b_i \\ \bar{a}}^\top G\matrix{c}{b_i \\ \bar{a}} ,\\
\mathrm{s.t.} & \hspace{.1in} \matrix{c}{b_i \\ \bar{a}}^\top \matrix{cc}{\frac{1}{N^2}\V_{\theta\theta} & 0 \\ 0 & \V_{xx}}\matrix{c}{b_i \\ \bar{a}} \nonumber \\ &\hspace{.3in}\leq 1-\frac{1}{N^2}\sum_{j\neq i} b_j^\top \V_{\theta\theta} b_j-\frac{1}{N^2}\sum_{j\neq i}V_{v_jv_j}.
\end{align}
\end{subequations}
If $\frac{1}{N^2}\sum_{j\neq i} b_j^\top \V_{\theta\theta} b_j+\frac{1}{N^2}\sum_{j\neq i}V_{v_jv_j}=1$, we know that $b_i^*=0$ and $\bar{a}^*=0$. Thus, we focus on the case where $\frac{1}{N^2}\sum_{j\neq i} b_j^\top \V_{\theta\theta} b_j+\frac{1}{N^2}\sum_{j\neq i}V_{v_jv_j}<1$. With change of variable 
$$
\matrix{c}{b'_i \\ \bar{a}'}\hspace{-.04in}=\hspace{-.04in}\frac{1}{\sqrt{1-\frac{1}{N^2}\sum_{j\neq i} b_j^\top \V_{\theta\theta} b_j -\frac{1}{N^2}\sum_{j\neq i}V_{v_jv_j}} }\hspace{-.03in}\matrix{c}{b_i \\ \bar{a}}\hspace{-.04in},
$$
we can rewrite~\eqref{eqn:appendix:optimization:2} as
\begin{equation*}
\begin{split}
\min_{\bar{a},b_i} & \hspace{.1in} \varrho \matrix{c}{b'_i \\ \bar{a}'}^\top G\matrix{c}{b'_i \\ \bar{a}'} ,\\
\mathrm{s.t.} & \hspace{.1in} \matrix{c}{b'_i \\ \bar{a}'}^\top \matrix{cc}{\frac{1}{N^2}\V_{\theta\theta} & 0 \\ 0 & \V_{xx}}\matrix{c}{b'_i \\ \bar{a}'} \leq 1,
\end{split}
\end{equation*}
where
$\varrho=1-\frac{1}{N^2}\sum_{j\neq i} b_j^\top \V_{\theta\theta} b_j-\frac{1}{N^2}\sum_{j\neq i} \V_{v_jv_j}$.
Following Lemma~\ref{lemma:eigenvector} in Appendix~\ref{App:UsefulLemma}, we know that the solution of this problem is indeed equal to
$$
\matrix{c}{b^{\prime *}_i \\ \bar{a}^{\prime *} }=\matrix{cc}{N\V_{\theta\theta}^{-1/2} & 0 \\ 0 & \V_{xx}^{-1/2}}\xi,
$$
where $\xi$ is the normalized eigenvector (i.e., $\|\xi\|_2=1$) corresponding to the smallest eigenvalue of the matrix
\begin{align*}
&\matrix{cc}{N\V_{\theta\theta}^{-1/2} & 0 \\ 0 & \V_{xx}^{-1/2}}G
\matrix{cc}{N\V_{\theta\theta}^{-1/2} & 0 \\ 0 & \V_{xx}^{-1/2}}\\ & \hspace{0.8in} =
\matrix{cc}{0 & -\V_{\theta\theta}^{1/2}\V_{xx}^{1/2} \\ -\V_{xx}^{1/2}\V_{\theta\theta}^{1/2} & -\V_{xx}}.
\end{align*}
This is indeed true because the above matrix always has at least one negative eigenvalue because its trace is equal to $\trace(-\V_{xx})$ which is negative (otherwise, the sensor would be better off by selecting $a_i=0$ and $b_i=0$). This solution implies that $\beta_i(\gamma_{-i})=(a_i^*,b_i^*,0)$ where
\begin{align*}
\matrix{c}{b_i^* \\ a_i^*}= &\sqrt{ 1-\frac{1}{N^2}\sum_{j\neq i} b_j^\top \V_{\theta\theta} b_j -\frac{1}{N^2}\sum_{j\neq i}V_{v_jv_j}}\nonumber\\ & \times\matrix{cc}{\hspace{-.04in} N\V_{\theta\theta}^{-1/2} \hspace{-.04in}& 0 \\ 0 & \hspace{-.04in}N\V_{xx}^{-1/2}\hspace{-.04in}}\xi \hspace{-0.03in}-\hspace{-.03in}\matrix{c}{0 \\ \hspace{-.06in}\sum_{j\neq i} a_j\hspace{-.06in}}.
\end{align*}
Now, one can check that the $\gamma^*=(a^*,b^*,0)$ with $a^*$ and $b^*$ defined as in the statement of the theorem is a fixed point of the best response mapping, that is, $\gamma^*=\beta_i((\gamma^*)_{j\neq i})$, $\forall i$.
\end{IEEEproof}

Because for the presented equilibrium in Theorem~\ref{tho:general}, $\xi_1$ is independent of $N$, the estimation quality in the receiver degrades with increasing $N$. This is because $a_i$ and $b_i$ are, respectively, decreasing and increasing functions of $N$, which means that each sensor puts more emphasis on its private information rather than the state of the system as $N$ grows. More precisely, we can show the following corollary.

\begin{corollary} \label{cor:multiple} Assume that $n_{y_i}=1$ for all $i\in\N$, $\V_{x\theta}=0$, and $\U_{\theta\theta}=0$. Let $(\hat{x}^*,(\gamma^*)_{i\in\N})$ be the equilibrium in affine strategies introduced in Theorem~\ref{tho:general}. Then, $\mathbb{E}\{\|x-\farhad{[\hat{x}^*((\gamma^*)_{i\in\N})]}((\gamma^*(x,\theta_i))_{i\in\N})\|_2^2\}$ is an increasing function of $N$, specifically, $\lim_{N\rightarrow \infty}\mathbb{E}\{\|x-\farhad{[\hat{x}^*((\gamma^*)_{i\in\N})]}((\gamma^*(x,\theta_i))_{i\in\N})\|_2^2\}=\trace(\V_{xx}$).
\end{corollary}

\begin{IEEEproof} \farhad{See Appendix~\ref{proof:cor:multiple}.}
%Please see~\cite{FarokhiCompleteManuscript2015} for a detailed proof.
\end{IEEEproof}

Corollary~\ref{cor:multiple} states a rather counter-intuitive result as it shows that, for the extracted affine equilibrium, the performance of the receiver (i.e., the quality of the estimate) degrades by summoning more sensors. This behavior can become even worse by expanding the policies of the sensors to also contain nonlinear mappings (see Remark~\ref{remark:linear_nonlinear}). By considering a slightly different model in the next section, we show that this result crucially depends on each sensor's belief about the others' strategic intention in equilibrium.

\begin{figure}
\centering
\includegraphics[width=1.0\columnwidth]{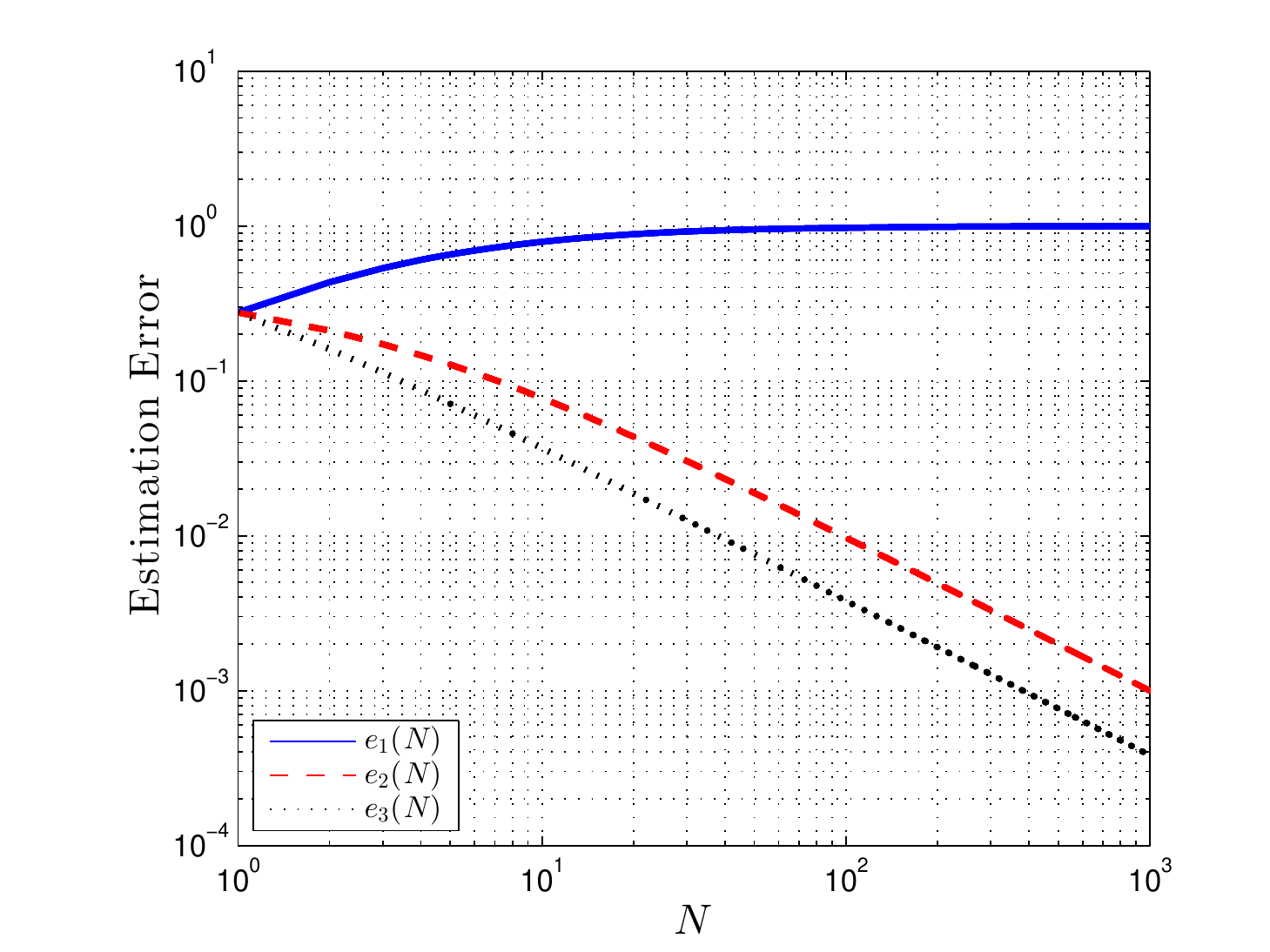}
\caption{Estimation error $\mathbb{E}\{\|x-\hat{x}(y)\|_2^2\}$ as a function of the number of sensors $N$ for different sensing scenarios.}
\label{fig:diagram1}
\end{figure}

\begin{example} \textsc{(Traffic Estimation):} Consider an example in which the receiver is interested in estimating the travel time on a single road. Note that this setup can be easily generalized to a neighborhood or a city by separately estimating the traffic on each road. In addition, considering the traffic flow conservation (i.e., the total inflow and outflow traffic are equal to each other in each junction), we can use the measurements from adjacent roads as side-channel information. The travel time is a scalar variable denoting the time that it takes to go from one end of the street to the other end, which varies according to the congestion level. At any given time of the day, using the historical data, we have a fairly accurate measurement of the average travel time. Therefore, the task at hand is to measure the innovation (i.e., the travel time minus its average) which is denoted by $x$. Although, in transportation literature, the travel times are assumed to follow a log-normal distribution for highways and urban areas, assuming a Gaussian distribution is also fairly common~\cite{noland2002travel} (note that, with a Gaussian distribution, the travel time may become negative with a nonzero probability, however, this probability will be negligible if the variance is small). The vehicles that drive along the road have an accurate measurement of $x$ by timing their trips. Now, imagine we have distributed a mobile application for crowd-sourcing estimation that asks the vehicles to register their message $y_i$, $i\in\N$, and, in return, it provides the community of its user with time-optimal trip planning. The application, as many of the available traffic applications, does not reveal the messages of the other vehicles (since it is rather useless for most users who simply wish to plan their trip). Clearly, in this example, we have $n_x=1$ and $n_{y_i}=1$ for all $i\in\N$. Furthermore, using an appropriate change of variable, we can always set $\V_{xx}=1$. Let us assume that $\V_{x\theta}=0$, which implies that the sensors preference does not depend on the actual state of the traffic (i.e., they under- or over-state the traffic irrespective of what is going on the road). Finally, let $\V_{\theta\theta}=1$. A key assumption is that the number of the participants $N$ needs to be fixed in advance and, more restrictively, to be known globally. In some cases, this quantity might be known \textit{a priori} as commercial crowd-sourcing applications tend to publicly advertise the number of the consented participants; however, a viable direction for future research could be to introduce individual beliefs on the number of participants for each sensor in order to avoid the dependency of the equilibrium to $N$. 

Following Theorem~\ref{tho:general}, under the described circumstances, we can calculate the sensors' signal
$$
y_i=\frac{0.8506}{\sqrt{0.7236+0.2763N}}x+\frac{0.5257N}{\sqrt{0.7236+0.2763N}}\theta_i.
$$
We also know that $\hat{x}^*(y)=\mathbb{E}\{x|(y_1+\cdots+y_N)/N\}$, which allows us to calculate the estimation error as a function of the number of sensors according to
\begin{align*}
e_1(N)&=\mathbb{E}\{\|x-\hat{x}^*(y)\|_2^2\}
=\frac{0.2763 N}{0.7236+0.2763 N}.
\end{align*}
Clearly, as $N$ grows, the quality of the estimation degrades. Figure~\ref{fig:diagram1} illustrates the estimation error variance $e_1(N)$ as a function of the number of sensors $N$ with a blue solid curve. As we can see, for large values of $N$, the crowd-sourcing technique does not provide any insight in the travel time.
\end{example}

\subsection{Static Estimation with Synchronous Herding Sensors} \label{sec:herding}
In this subsection, we consider ``herding" equilibria in which sensors imitate each other.  The herding scenario models an interesting intermediate situation where each sensor is refined enough to recognize that others may also be strategically misreporting but, having limited cognitive abilities, assumes that they simply copy its own policy.

Note that, because of the symmetry  assumptions, $\Gamma_i=\Gamma_j$ for all $i,j\in\N$ and, hence, we use $\Gamma$ to denote these sets.

\begin{definition}\textsc{(Herding Equilibrium in Affine Strategies):} A pair $(\hat{x}^*,\gamma^*)\in\farhad{\mathcal{C}(\Gamma,\Upsilon)}\times \Gamma$ constitutes a herding equilibrium in affine strategies if 
\begin{subequations}
\begin{align}
\hat{x}^* &\in\argmin_{\hat{x}\in\farhad{\mathcal{C}(\Gamma,\Upsilon)}} \mathbb{E}\{\|x-\farhad{[\hat{x}(\gamma^*)]}((\gamma^*(x,\theta_j))_{j\in\N})\|_2^2\}, \label{eqn:optimality:herding} \\ 
\gamma^* &\in \argmin_{\gamma\in\Gamma} \mathbb{E}\{\|(x+\theta_i)-\farhad{[\hat{x}^*(\gamma)]}((\gamma(x,\theta_j))_{j\in\N})\|_2^2\},
\end{align}
\end{subequations}
for all $i\in\N$.
\end{definition}

In this definition, all the sensors must deviate at the same time whereas, in Definition~\ref{def:generalequi}, the sensors could deviate unilaterally. Therefore, a herding equilibrium does not constitute an equilibrium in the sense of Definition~\ref{def:generalequi} since one of the sensors might benefit from breaking away from the herd; i.e., by not employing the same strategy as the other sensors. 

\begin{lemma} \label{lemma:6} If $\gamma_i=(a,b,\V_{vv})$ for all $i\in\N$, then
\begin{equation*}
\begin{split}
\mathbb{E}\{\|(x+\theta_i)-\mathbb{E}\{x|y\}\|_2^2\}=&\;\mathbb{E}\{\|(x+\bar{\theta})-\mathbb{E}\{x|y\}\|_2^2\}\\&+\frac{N-1}{N}\trace\left(\V_{\theta\theta}-\U_{\theta\theta}\right)
\end{split}
\end{equation*}
where $\bar{\theta}=(\theta_1+\cdots+\theta_N)/N$.
\end{lemma}

\begin{IEEEproof} See Appendix~\ref{proof:lemma:6}.
\end{IEEEproof}

Lemma~\ref{lemma:6} shows that when the sensors herd, we can replace them with a single sensor with private information $(\theta_1+\cdots+\theta_N)/N$. Now, intuitively, because of the Law of Large Numbers, one might expect that, as $N$ grows, the agents' contributions cancel each other and, eventually, the receiver may have access to the perfect estimation. We show this in the rest of the subsection. 

\begin{theorem} \label{tho:herding} Assume that $n_{y_i}=1$ for all $i\in\N$, $\V_{x\theta}=0$, and $\U_{\theta\theta}=0$. There exists a herding equilibrium in affine strategies in which the receiver follows $\farhad{[\hat{x}^*(\gamma)]}(y)=\mathbb{E}\{x|(y_1+\dots+y_N)/N\}$ and sensor $S_i$, $i\in\N$, employs the linear policy $\gamma^*=(a^*,b^*,0)$ where
\begin{equation*}
\begin{split}
\matrix{c}{b^*\\ a^*}=\matrix{cc}{\sqrt{N}\V_{\theta\theta}^{-1/2} & 0 \\ 0 & \V_{xx}^{-1/2}}\zeta,
\end{split}
\end{equation*}
and $\zeta$ is the normalized eigenvector (i.e., $\|\zeta\|_2=1$) corresponding to the smallest eigenvalue of the matrix
\begin{align*}
\matrix{cc}{0 & -\frac{1}{\sqrt{N}}\V_{\theta\theta}^{1/2}\V_{xx}^{1/2} \\ -\frac{1}{\sqrt{N}}\V_{xx}^{1/2}\V_{\theta\theta}^{1/2} & -\V_{xx}}.
\end{align*}
Furthermore, the sensors' policy of the form $\kappa\gamma^*$, for some $\kappa\in \mathbb{R}\setminus\{0\}$, along side the receiver's policy $\hat{x}^*$, also constitutes a herding equilibrium in affine strategies.
\end{theorem}

\begin{IEEEproof} Following the result of Lemma~\ref{lemma:1}, we know that the receiver cannot improve its estimation error by following different strategy. Following Lemma~\ref{lemma:6}, all the sensors would have the same cost which is equal to the cost function an aggregate sensor with private information $\bar{\theta}=(\theta_1+\cdots+\theta_N)/N$ (up to a constant term). Therefore, in this scenario, we can replace all the sensors with a single sensor. Doing so, we can use Theorem~\ref{tho:1} to calculate the equilibrium for the case that the side channel information is ignored. In this case, we can see that the best response of the sensor can be extracted from optimization problem
\begin{align*}
(\V_{xz}^*,\V_{\bar{\theta} z}^*)\in\argmin_{\V_{xz},\V_{\bar{\theta} z}} & \matrix{c}{\V_{xz} \\ V_{\bar{\theta} z}}^{\hspace{-.03in}\top} \hspace{-.05in}\matrix{cc}{-I & -I \\ -I & 0} \hspace{-.05in}\matrix{c}{\V_{xz} \\ V_{\bar{\theta} z}}, \\
\mathrm{s.t.} \hspace{.15in} & \matrix{c}{\V_{xz} \\ V_{\bar{\theta} z}}^{\hspace{-.03in}\top} \hspace{-.05in}\matrix{cc}{\V_{xx}^{-1} & 0 \\ 0 & \V_{\bar{\theta}\bar{\theta}}^{-1}} \hspace{-.05in}\matrix{c}{\V_{xz} \\ V_{\bar{\theta} z}}\hspace{-.03in}\leq\hspace{-.03in} 1.
\end{align*}
Again, using Lemma~\ref{lemma:eigenvector}, we can deduce that
$$
\matrix{c}{\V_{xz}^* \\ V_{\bar{\theta} z}^*}=\matrix{cc}{\V_{xx}^{1/2} & 0 \\ 0 & \V_{\bar{\theta}\bar{\theta}}^{1/2}}\zeta,
$$
where $\zeta$ is the normalized eigenvector (i.e., $\|\zeta\|_2=1$) of the smallest eigenvalue of 
\begin{align*}
&\matrix{cc}{\V_{xx}^{1/2} & 0 \\ 0 & \V_{\bar{\theta}\bar{\theta}}^{1/2}} \matrix{cc}{-I & -I \\ -I & 0}\matrix{cc}{\V_{xx}^{1/2} & 0 \\ 0 & \V_{\bar{\theta}\bar{\theta}}^{1/2}}\\& \hspace{1in} =\matrix{cc}{-\V_{xx} & -\frac{1}{\sqrt{N}} \V_{xx}^{1/2}\V_{\theta\theta}^{1/2} \\ \frac{1}{\sqrt{N}}\V_{\theta\theta}^{1/2}\V_{xx}^{1/2} & 0},
\end{align*}
where the second equality is a direct consequence of the fact that $\V_{\bar{\theta}\bar{\theta}}=N^{-1}\V_{\theta\theta}$. Now, following the same argument as in the proof of Theorem~\ref{tho:1}, we can extract the explicit solution presented in the statement of the theorem.
\end{IEEEproof}

\farhad{\begin{remark} Note that the results of Theorem~\ref{tho:herding} holds for general mappings and not the set of affine policies. This is because, in the proof of this theorem, we refer to the results of Theorem~\ref{tho:1} which holds for more general policies.
\end{remark}}

\begin{corollary} \label{cor:multiple:herding} Assume that $n_{y_i}=1$ for all $i\in\N$, $\V_{x\theta}=0$, $\U_{\theta\theta}=0$, $\V_{xx}=\eta_x I$, and $\V_{\theta\theta}=\eta_\theta I$. Let $(\hat{x}^*,(\gamma^*)_{i\in\N})$ be the herding equilibrium in affine strategies introduced in Theorem~\ref{tho:herding}. Then, $\mathbb{E}\{\|x-\farhad{[\hat{x}^*(\gamma^*)]}((\gamma^*(x,\theta_i))_{i\in\N})\|_2^2\}$ is a decreasing function of $N$. Furthermore, $\lim_{N\rightarrow \infty}\mathbb{E}\{\|x-\farhad{[\hat{x}^*(\gamma^*)]}((\gamma^*(x,\theta_i))_{i\in\N})\|_2^2\}=0$.
\end{corollary}

\begin{IEEEproof} \farhad{See Appendix~\ref{proof:cor:multiple:herding}.}
%See~\cite{FarokhiCompleteManuscript2015} for a detailed proof.
\end{IEEEproof}

%\farhad{
%\begin{remark} Although the herding can be motivated by the bounded rationality of the players, we may be able to design appropriate estimators at the receiver that sacrifice ``optimality'', in the sense of~\eqref{eqn:optimality:herding}, to induce herding even when dealing with fully rational players. This can be done, for instance, by rejecting outlier messages. According to Corollary~\ref{cor:multiple:herding}, such a design is clearly beneficial as it promotes a ``good behaviour'' from the perspective of the receiver. This avenue was recently explored in~\cite{7058370}. 
%\end{remark}
%}

\setcounter{example}{1}
\begin{example} \textsc{(Traffic Estimation, Cont'd):} Following Theorem~\ref{tho:herding}, we can calculate the sensors' signal in the herding equilibrium as
\begin{align*}
y_i=\frac{\sqrt{2}}{\sqrt{N\hspace{-.03in}-\hspace{-.03in}\sqrt{N(N+4)}+4}}x+\frac{\sqrt{2N}(\sqrt{N (N + 4)}\hspace{-.03in}-\hspace{-.03in}N)}{\sqrt{N\hspace{-.03in}-\hspace{-.03in}\sqrt{N(N+4)}+4}} \theta_i
\end{align*}
Furthermore, we can calculate the estimation error in the receiver as
$$
e_2(N)=1-\frac{2}{N+4-\sqrt{N(N+4)}}.
$$
In this case, it is evident that the quality of the estimation improves as the number of sensors grows which demonstrates why herding between strategic sensors is a virtue. 

Now, let us consider a rival scenario in which the sensors are not strategic, however, they have access to the noisy state measurements. Therefore, they transmit $y'_i=x+u_i$ where $(u_i)_{i\in\N}$ are i.i.d Gaussian random variables so that $\mathbb{E}\{u_i\}=0$ and $\mathbb{E}\{u_i^2\}=\sigma$ for all $i\in\N$. We also assume that $\mathbb{E}\{u_i x\}=0$ for all $i\in\N$.  The estimation error is $e_3(N)=\sigma/(\sigma+N)$. 

Figure~\ref{fig:diagram1} illustrates the estimation error variances $e_2(N)$ and $e_3(N)$ as a function of the number of sensors $N$ to visually compare the estimation error of different scenarios with each other. Here, we have set $\sigma=0.3820$ so that all the schemes have equal error at $N=1$. Interestingly, we can note that
$$
\lim_{N\rightarrow \infty} \frac{e_2(N)}{e_3(N)}=\frac{1}{\sigma}.
$$
Hence, if $\sigma>1$, employing many strategic but accurate sensors that herd is better than employing many honest but noisy sensors.
\end{example}

\subsection{Static Estimation with Asynchronous Independent Sensors}
\label{sec:asynchrnuous}
In many crowd-sensing applications, the users enter their data sequentially (and not simultaneously). This is the case, firstly, because they do not coordinate their actions and, secondly, because they are not measuring the \farhad{state} of nature at the same time. This creates an interest for investigating estimation in the presence of strategic sensor with asynchronous communication structure, which is the topic of this subsection. We still assume that the sensors and the receiver are connected to each other using the communication network presented in Figure~\ref{fig:1}; however, the sensors transmit their signals sequentially and the receiver computes an estimate after each transmission. At time step $i\in\N$, only sensor $S_i$ transmits the message $y_i\in\mathbb{R}^{n_{y_i}}$ to the receiver $R$. Note that the definition of ``time step'', in this subsection, is not the same as in Subsection~\ref{sec:singlesensor:dynamic}. Here, the evolution of ``time'' merely points out the order of the sensors and the underlying estimation problem is still static. We assume that the sensors have access to all the previously transmitted messages (and may consider them as side-channel information), however, as we will see later, at the equilibria, the sensor do not use this information. Similar to the previous subsections, each sensor $S_i$ has access to the exact measurement of the state $x\in\mathbb{R}^{n_x}$ and its own private parameter $\theta_i\in\mathbb{R}^{n_x}$.

At time step $i\in\N$, the receiver $R$ computes the optimal estimate by \farhad{minimizing $\mathbb{E}\{\|x-\farhad{\upsilon_i}((y_k)_{k=1}^{i})\|_2^2\}$ over $\Upsilon_i$ denoting} the set of all Lebesgue-measurable functions from $\prod_{k=1}^{i}\mathbb{R}^{n_{y_k}}$ to $\mathbb{R}^{n_x}$. The goal of the sensor $S_i$, $i\in\N$, is to \farhad{minimize $\mathbb{E}\{\|(x+\theta)-\farhad{\upsilon_i}((y_k)_{k=1}^{i-1},\gamma_i(x,(y_k)_{k=1}^{i-1},\theta_i))\|_2^2\}$ over $\Gamma_i$, which} is the set of appropriate stochastic mappings defined in the same fashion as in Subsections~\ref{sec:singlesensor:static} and~\ref{sec:singlesensor:dynamic}. Unlike the previous subsections, we do not restrict ourselves to affine policies.

\begin{definition}\textsc{(Equilibrium with Asynchronous Communication):} \label{def:Asynchrnuous} A tuple $((\hat{x}^*_i)_{i\in\N},\linebreak(\gamma_i^*)_{i\in\N}) \in\prod_{i\in\N}\farhad{\mathcal{C}(\Gamma_i,\Upsilon_i)}\times\prod_{i\in\N}\Gamma_i$ constitutes an equilibrium with asynchronous communication if for all $i\in\N$,
\begin{subequations}
\begin{align}
\hat{x}^*_i &\in \argmin_{\hat{x}_i\in\farhad{\mathcal{C}(\Gamma_i,\Upsilon_i)}} \mathbb{E}\{\|x-\farhad{[\hat{x}_i(\gamma_i^*)]}((y_k^*)_{k=1}^{i})\|_2^2\}, \\ 
\gamma_i^* &\in \argmin_{\gamma_i\in\Gamma_i} \mathbb{E}\{\|(x+\theta)-\farhad{[\hat{x}_i^* (\gamma_i)]}\nonumber\\&\hspace{.9in}((y_k^*)_{k=1}^{i-1},\gamma_i(x,(y_k^*)_{k=1}^{i-1},\theta_i))\|_2^2\},
\end{align}
\end{subequations}
where $y^*_k=\gamma_k^*(x,(y_{k'}^*)_{k'=1}^{k-1},\theta_k)$ for all $k\in\N$.
\end{definition}

\begin{remark} Here, the sensor are myopic, i.e., the  cost of each sensor is only a function of $\farhad{[\hat{x}_i^*(\gamma_i)]}((y_k^*)_{k=1}^{i-1},\gamma_i(x,(y_k^*)_{k=1}^{i-1},\theta_i))$, which is the estimate after its transmission and not the estimate at the end (after all the transmissions). This is a particularly useful concept if the sensors do not know the number of active participants in the estimation scheme (but they can observe, at least, the number of the sensor that have already contributed).
\end{remark}

\begin{theorem} \label{tho:2} 
There exists an equilibrium in which, at step $i\in\N$, the receiver uses the LMS estimator
\begin{align} \label{eqn:recieverbestresponse:async}
\farhad{[\hat{x}^*_i(\gamma_i)]}((y^*_{k})_{k=1}^{i})=\mathbb{E}\{x|y_1^*,\dots,y_i^*\},
\end{align}
while the sensor $S_i$, $i\in\N$, uses the policy
\begin{align} \label{eqn:sensorbestresponse:async}
y^*_i
&=\gamma^*_i(x,(y_k^*)_{k=1}^{i-1},\theta)\nonumber\\
&=\alpha_1^\top x+\alpha_2^\top \theta+v_i.
\end{align}
In the sensor's policy, $v\in\mathbb{R}^{n_{y_i}}$ is a Gaussian random variable with zero mean and covariance matrices
\begin{equation*}
\matrix{ccc}{\V_{vx} & \V_{v\theta} & \V_{vy}}=\matrix{ccc}{0 & 0 &0 },
\end{equation*}
\begin{equation*} 
V_{vv}=I-\matrix{c}{\alpha_1 \\ \alpha_2}^\top \Psi_i \matrix{c}{\alpha_1 \\ \alpha_2},
\end{equation*}
in which 
\begin{align*} 
\matrix{c}{\alpha_1 \\ \alpha_2}\in\argmin_{\xi\in\mathbb{R}^{2n_x\times n_{y_i}}} & \trace\left({\xi'}^\top \Psi_i \matrix{cc}{-I & -I \\ -I & 0} \Psi_i \xi'\right),\\
\mathrm{s.t.} \hspace{.18in}& \,\,{\xi'}^\top \Psi_i {\xi'}\leq I,
\end{align*}
with 
\begin{align*}
\Psi_i=\matrix{ccc}{\V_{xx} & \V_{x\theta} \\ \V_{\theta x} & \V_{\theta\theta}}-\matrix{c}{\V_{x\psi_i} \\ \V_{\theta \psi_i}}\V_{\psi_i \psi_i}^{-1}\matrix{c}{\V_{x\psi_i} \\ \V_{\theta \psi_i}}^{\top},
\end{align*}
$$
\psi_i=\matrix{ccc}{y^{*\top}_1 & \cdots & y^{*\top}_{i-1}}^\top.
$$
Furthermore, the sensor's policy of the form $(\kappa_i\gamma_i^*)_{i\in\N}$, for some $\kappa_i\in \mathbb{R}\setminus\{0\},\forall i\in\N$, along side the receiver's policy $(\hat{x}_i^*)_{i\in\N}$, also constitutes an equilibrium. All these equilibria result in the same estimation error variance at the receiver.
\end{theorem}

\begin{IEEEproof} The proof follows from utilizing Theorem~\ref{tho:1} sequentially and  treating all accumulated
information at this time $y_1^*,\dots,y_{i-1}^*$ as the side-channel information.
\end{IEEEproof}

%\begin{remark} In this case, we do not need to assume that all the sensors are strategic. If a specific nonstrategic sensor has access to the noisy-measurements of the state, it simply transmit it to the receiver, that is, $y^*_i$ needs not follow~\eqref{eqn:sensorbestresponse:async} and it can be equal to the information that is available to that sensor. In this case, however, we should assume that this specific sensor does not have access to the precise measurement of the state since, otherwise, the problem becomes trivial.
%\end{remark}

\section{Conclusions and Future Work} \label{sec:conclusion}
In this paper, we investigated static and dynamic estimation with strategic self-interested sensors using a game theoretic viewpoint. We first calculated an equilibrium for the single sensor case in the presence of an honest but noisy side-channel information for both static and dynamic estimation problems. Interestingly, the sensor's policy turned out to be memory-less in the dynamic case for the characterized equilibrium. Then, we extended the setup to study static estimation with multiple sensors when (\textit{i}) the sensors are strategic but restricted to using affine policies and when (\textit{ii}) they herd (i.e., they imitate each others' policies). We showed that when the sensors are herding, the receiver can indeed estimate the state of the system with a large number of sensors which does not seem to be possible in the other case. Finally, we partly extended the results to the case in which the sensors communicate sequentially. \farhad{An avenue for future research is to remove the i.i.d. assumption from the underlying random variables. Further f}uture work can focus on using mechanism design theory to appropriately incentivize the sensors to communicate truthfully, which would allow us to better understand the price of information in networked estimation. 

\bibliographystyle{ieeetr}
\bibliography{compile_new}

\newpage

\appendices

\section{} \label{App:UsefulLemma}
\begin{lemma} \label{lemma:eigenvector} For any $n\in\mathbb{N}$, let $X\in\mathbb{R}^{n\times n}$ be a symmetric matrix with at least one negative eigenvalue. Moreover, let $\xi\in\mathbb{R}^n$ denote the normalized eigenvector corresponding to the smallest eigenvalue of $X$. Then
\begin{align*}
\xi\in\argmin_{x\in\mathbb{R}^n} &\, x^\top X x,\\
\mathrm{s.t.} \hspace{.15in} & \, x^\top x\leq 1.
\end{align*}
\end{lemma}

\begin{IEEEproof}
To prove this lemma, we first show that the inequality constraint $x^\top x\leq 1$ can be replaced by the equality constraint $x^\top x=1$ without changing the optimal solution (i.e., the constraint is active for all the optimal solutions). Then, the rest of the proof automatically follows from the Courant--Fischer--Weyl min-max principle~\cite[p.\,58]{bhatia1997matrix}. To prove this, note that the optimization problem admits, at least, one solution because the feasible set is a compact set (i.e., it is closed and bounded subset of $\mathbb{R}^n$) and the cost function is continuous. Pick any optimal solution $\bar{x}$. We show that $\bar{x}^\top \bar{x}=1$. Assume that this not the case, that is, $\bar{x}^\top \bar{x}=\delta<1$. Let $\zeta\in\mathbb{R}^n$ be the normalized eigenvector corresponding to the negative eigenvalue in the statement of the lemma. Now, we define $\tilde{x}=(1/\sqrt{\delta})\bar{x}$. By definition, $\tilde{x}^\top \tilde{x}=\bar{x}^\top \bar{x}/\delta=1$. Therefore, we get $\tilde{x}^\top X\tilde{x}=\frac{1}{\delta} \bar{x}^\top X\bar{x}<\bar{x}^\top X\bar{x},$ where the strict inequality follows from that $\delta<1$  and $\bar{x}^\top X\bar{x}\leq \zeta^\top X\zeta<0$. This is in contradiction with $\bar{x}$ being an optimal solution. 
\end{IEEEproof}

\farhad{
\section{Proof of Proposition~\ref{prop:1}} 
\label{proof:prop:1}
Notice that we can rewrite $\Xi'$ as $\Omega^\top\Omega$ where
\begin{align*}
\Omega=
\begin{bmatrix}
\Omega_{11} & \Omega_{12} \\
0 & \Omega_{22}
\end{bmatrix}=
\begin{bmatrix}
\tilde{V}_{xx}^{1/2} & \tilde{V}_{xx}^{-1/2}\tilde{V}_{x\theta} \\
0 & (\tilde{V}_{\theta\theta}-\tilde{V}_{\theta x}\tilde{V}_{xx}^{-1}\tilde{V}_{x\theta})^{1/2}
\end{bmatrix}
\end{align*}
with $\tilde{V}_{xx}=V_{xx}-V_{xy}V_{yy}^{-1}V_{yx}$, $\tilde{V}_{x\theta}=V_{x\theta}-V_{xy}V_{yy}^{-1}V_{y\theta}$, and $\tilde{V}_{\theta\theta}=V_{\theta\theta}-V_{\theta y}V_{yy}^{-1}V_{y\theta}$. Using Schur complement on $\Xi'>0$, we can see that $\tilde{V}_{xx}>0$ and $\tilde{V}_{\theta\theta}-\tilde{V}_{\theta x}\tilde{V}_{xx}^{-1}\tilde{V}_{x\theta}>0$. Now, by introducing the change of variable $\eta=\Omega \xi'$, we may transform~\eqref{eqn:cor:notafunctionofy:optimzation} into
\begin{equation} \label{eqn:nontrivial:1}
\begin{split}
\min_{\eta\in\mathbb{R}^{2n_x}} \,\,& \eta^\top \Omega\matrix{cc}{-I & -I \\ -I & 0} \Omega^\top\eta,\\
\mathrm{s.t.} \,\,\,\,\,& \eta^\top \eta\leq 1.
\end{split}
\end{equation}
Following the same line of reasoning as in the proof of Corollary~\ref{cor:1}, we can see that the solution of~\eqref{eqn:nontrivial:1}, denoted by $\eta^*$, is the normalized eigenvector corresponding to the smallest eigenvalue of 
\begin{align*}
&\Omega\matrix{cc}{-I & -I \\ -I & 0} \Omega^\top
\\&\hspace{.4in}=\matrix{cc}{-\Omega_{11}\Omega_{11}^\top-\Omega_{11}\Omega_{12}^\top-\Omega_{12}\Omega_{11}^\top &-\Omega_{11}\Omega_{22}^\top \\ -\Omega_{22}\Omega_{11}^\top & 0}.
\end{align*}
We show that, for this solution, $\alpha_1\neq 0$ by \textit{reductio ad absurdum}. To do so, assume that $\alpha_1=0$. Therefore, 
$$
\eta^*=\Omega \matrix{c}{0 \\ \alpha_2}=
\matrix{c}{\Omega_{12} \\ \Omega_{22}}\alpha_2
$$
Hence, there should exists $\lambda<0$
(by same argument as in the proof of Corollary~\ref{cor:1}) such that
\begin{align*}
&\matrix{cc}{-\Omega_{11}\Omega_{11}^\top-\Omega_{11}\Omega_{12}^\top-\Omega_{12}\Omega_{11}^\top &-\Omega_{11}\Omega_{22}^\top \\ -\Omega_{22}\Omega_{11}^\top & 0}\matrix{c}{\Omega_{12} \\ \Omega_{22}}\alpha_2\\
&=
\matrix{c}{-\tilde{V}_{xx}^{1/2}\tilde{V}_{x\theta}-\tilde{V}_{xx}^{1/2}\tilde{V}_{\theta\theta}-\Omega_{12}V_{x\theta} \\ -\Omega_{22}V_{x\theta}}\alpha_2
=\lambda \matrix{c}{\Omega_{12} \\ \Omega_{22}}\alpha_2.
\end{align*}
Noting that $\Omega_{22}>0$, we get
\begin{subequations}
\begin{align}
(-\tilde{V}_{xx}^{1/2}\tilde{V}_{x\theta}-\tilde{V}_{xx}^{1/2}\tilde{V}_{\theta\theta}-\Omega_{12}V_{x\theta})\alpha_2&=\lambda\Omega_{12}\alpha_2,\label{eqn:Omega:sub1}
\\
-V_{x\theta}\alpha_2&=\lambda\alpha_2. \label{eqn:Omega:sub2}
\end{align}
\end{subequations}
Substituting~\eqref{eqn:Omega:sub2} into~\eqref{eqn:Omega:sub1} while noting that $\tilde{V}_{xx}^{1/2}>0$ gives
\begin{align*}
(-\tilde{V}_{x\theta}-\tilde{V}_{\theta\theta})\alpha_2=0,
\end{align*}
or, equivalently,
\begin{align*}
\tilde{V}_{\theta\theta}\alpha_2=
-\tilde{V}_{x\theta}\alpha_2=
\lambda\alpha_2,
\end{align*}
which is in contradiction with the fact that $\tilde{V}_{\theta\theta}>0$. This proves that $\alpha_1\neq 0$. The proof of that $\alpha_2\neq 0$ follows a similar argument.}

\section{Proof of Lemma~\ref{lemma:1}}\label{proof:lemma:1}
Let us define the random variables $\bar{y}=(y_1+\dots+y_N)/N$ and $\tilde{y}_i=y_i-\bar{y}$ for $i\in\N$. Moreover, let $y=(y_i)_{i\in\N}$. It is evident that no piece of information is lost with this change of variable because $\spans((y_i)_{i\in\N})=\spans((\tilde{y}_i)_{i\in\N})\oplus \spans(\bar{y}).$  Therefore, we have $\hat{x}(y)=\mathbb{E}\{x|y_1,\dots,y_N\}=\mathbb{E}\{x|\bar{y},\tilde{y}_1,\dots,\tilde{y}_N\}$, and as a result,
\begin{align} 
\mathbb{E}\{x|y\}=&\matrix{cccc}{\V_{x\bar{y}} & \V_{x\tilde{y}_1} & \cdots & \V_{x\tilde{y}_N}}\nonumber \\& \times\hspace{-.05in}\matrix{cccc}{\V_{\bar{y}\bar{y}} & \V_{\bar{y}\tilde{y}_1} & \cdots & \V_{\bar{y}\tilde{y}_N} \\ \V_{\tilde{y}_1\bar{y}} & \V_{\tilde{y}_1\tilde{y}_1} & \cdots & \V_{\tilde{y}_1\tilde{y}_N} \\ \vdots & \vdots & \ddots & \vdots \\ \V_{\tilde{y}_N\bar{y}} & \V_{\tilde{y}_N\tilde{y}_1} & \cdots & \V_{\tilde{y}_N\tilde{y}_N}}^{\hspace{-.03in}-1}\hspace{-.07in}\matrix{c}{\hspace{-.05in}\bar{y}\hspace{-.05in}\\ \hspace{-.05in}\tilde{y}_1 \hspace{-.05in}\\ \hspace{-.05in}\vdots \hspace{-.05in}\\ \hspace{-.05in}\tilde{y}_N\hspace{-.05in}}\hspace{-.04in}.\label{eqn:2}
\end{align}
Now, we can easily show that
\begin{equation*}
\begin{split}
\V_{\tilde{y}_i\bar{y}}
&=\mathbb{E}\Big\{\Big(b^\top \theta_i +v_i-\frac{1}{N}\sum_{j=1}^N (b^\top \theta_j +v_j)\Big)\\[-.7em]& \hspace{.8in}\times \Big(a^\top x+\frac{1}{N}\sum_{j=1}^N (b^\top \theta_j +v_j) \Big)^\top\Big\}
\\&=b^\top V_{\theta x}a+\frac{1}{N} b^\top V_{\theta\theta} b+\frac{N-1}{N} b^\top\U_{\theta\theta}b \\[-.3em]&\;\;\;\;-\frac{1}{N}\sum_{j=1}^N b^\top\V_{\theta x}a-\frac{1}{N^2} \sum_{j=1}^N\sum_{k=1}^N b^\top \mathbb{E}\{\theta_j\theta_k^\top\}b\\[-.5em]&\;\;\;\;+\frac{1}{N}\V_{vv}-\frac{1}{N^2}\sum_{j=1}^N \V_{vv} 
=0,
\end{split}
\end{equation*}
and
\begin{equation*}
\begin{split}
\V_{x\tilde{y}_i}
&=\mathbb{E}\Big\{x\Big(b^\top \theta_i +v_i-\frac{1}{N}\sum_{j=1}^N (b^\top \theta_j +v_j)\Big)^\top\Big\}
\\[-.5em]&=\V_{x\theta} b-\frac{1}{N} \sum_{j=1}^N \V_{x\theta}b=0.
\end{split}
\end{equation*}
Substituting these identities inside~\eqref{eqn:2} results in
\begin{equation*}
\begin{split}
\mathbb{E}\{x|y\}&=\matrix{cccc}{\V_{x\bar{y}} & 0 & \cdots & 0} \\&\;\;\;\;\times\matrix{cccc}{\V_{\bar{y}\bar{y}} & 0 & \cdots & 0 \\ 0 & \V_{\tilde{y}_1\tilde{y}_1} & \cdots & \V_{\tilde{y}_1\tilde{y}_N} \\ \vdots & \vdots & \ddots & \vdots \\ 0 & \V_{\tilde{y}_N\tilde{y}_1} & \cdots & \V_{\tilde{y}_N\tilde{y}_N}}^{-1}\matrix{c}{\bar{y}\\ \tilde{y}_1 \\ \vdots \\ \tilde{y}_N}\\&= \V_{x\bar{y}}\V_{\bar{y}\bar{y}}^{-1}\bar{y}\\&=\mathbb{E}\{x\,|\,\bar{y}\}.
\end{split}
\end{equation*}
This concludes the proof.

\section{Proof of Corollary~\ref{cor:multiple}}
\label{proof:cor:multiple}
First, note that
\begin{align*}
\mathbb{E}\{\|x-&\farhad{[\hat{x}^*((\gamma^*)_{i\in\N})]}((\gamma^*(x,\theta_i))_{i\in\N})\|_2^2\}\\
&=\mathbb{E}\{\|x-\mathbb{E}\{x|\bar{y}\}\|_2^2\}\\
&=\trace(\V_{xx}-\V_{x\bar{y}}\V_{\bar{y}\bar{y}}^{-1}\V_{\bar{y}x})\\
&=\trace(\V_{xx}-\V_{xx} a^* {a^*}^\top \V_{xx})\\
&=\trace(\V_{xx})-\frac{\trace(\V_{xx}^{1/2}\xi_2 \xi_2^\top \V_{xx}^{1/2})}{1+(N-1)\xi_1^\top \xi_1}.
\end{align*}
Now, we prove that $\xi_1\neq 0$ using \textit{reductio ad absurdum}. To do so, assume that $\xi_1=0$. Clearly, $\xi_2\neq 0$ in that case (since $\xi$ is an eigenvector). We have
\begin{align*}
&\matrix{cc}{
0 & -\V_{\theta\theta}^{1/2}\V_{xx}^{1/2} \\ 
-\V_{xx}^{1/2}\V_{\theta\theta}^{1/2} & -\V_{xx}}
\matrix{c}{0 \\ \xi_2}
\\&\hspace{1.5in}=\matrix{c}{-\V_{\theta\theta}^{1/2}\V_{xx}^{1/2} \xi_2 \\ -\V_{xx}\xi_2},
\end{align*}
which is valid (because $\xi$ is an eigenvector) only if $-\V_{\theta\theta}^{1/2}\V_{xx}^{1/2} \xi_2=0$. This leads to a contradiction since $\xi_2\neq 0$. Following the same line of reasoning, we can also prove that $\xi_2\neq 0$. Now, notice that $\xi_1,\xi_2$ do not depend on $N$ (because the matrix for which they serve as an eigenvector is not a function of $N$). This concludes the proof.

\section{Proof of Lemma~\ref{lemma:6}} \label{proof:lemma:6}
First, notice that
\begin{align} 
\mathbb{E}\{\|(x+&\theta_i)-\mathbb{E}\{x|y\}\|_2^2\}\nonumber\\
=&\trace(\mathbb{E}\{(\theta_i+x)(\theta_i+x)^\top-(x+\theta_i)\mathbb{E}\{x|y\}^\top\nonumber\\
&-\mathbb{E}\{x|y\}(\theta_i+x)^\top+ \mathbb{E}\{x|y\}\mathbb{E}\{x|y\}^\top\}).
\label{eqn:exansion:1}
\end{align}
Following Lemma~\ref{lemma:1}, we know that the optimal estimate $\mathbb{E}\{x|y\}$ is of the following form
\begin{equation*}
\begin{split}
\mathbb{E}\{x|y\}&=K\bar{y}\\&=Ka^\top x+Kb^\top \bar{\theta}+K\bar{v},
\end{split}
\end{equation*}
for some appropriately selected constant $K\in\mathbb{R}^{n_x}$. Here, $\bar{y}=N^{-1}\sum_{j\neq i} y_j$, $\bar{\theta}=N^{-1}\sum_{j=1}^N \theta_j$, and $\bar{v}=N^{-1}\sum_{j=1}^N v_j$.  
\begin{figure*}
\begin{align}
\mathbb{E}\{(x+\theta_i)\mathbb{E}\{x|y\}^\top\}
&=\mathbb{E}\Big\{(x+\theta_i)x^\top aK^\top + \frac{1}{N}\sum_{j=1}^N x\theta_j^\top bK^\top + \frac{1}{N}\sum_{j=1}^N \theta_i\theta_j^\top bK^\top +\frac{1}{N}\sum_{j=1}^N (x+\theta_i)v_j^\top K^\top\Big\}\nonumber\\[-.5em]
&=\mathbb{E}\Big\{\left(x+\bar{\theta}\right)x^\top aK^\top + \frac{1}{N}\sum_{j=1}^N x\theta_j^\top bK^\top + \frac{1}{N^2}\sum_{j=1}^N\sum_{t=1}^N \theta_t\theta_j^\top bK^\top +\frac{1}{N}\sum_{j=1}^N \left(x+\bar{\theta}\right)v_j^\top K^\top\Big\} \nonumber\\[-.5em]
&=\mathbb{E}\left\{\left(x+\bar{\theta}\right)\left[Ka^\top x+ Kb^\top \bar{\theta}+K\bar{v} \right]^\top\right\} \nonumber\\[-.5em]
&=\mathbb{E}\left\{\left(x+\bar{\theta}\right)\mathbb{E}\{x|y\}^\top\right\},
\label{eqn:exansion:2}
\end{align}
\hrulefill
\end{figure*}
Therefore, we can prove the identity in~\eqref{eqn:exansion:2} in which the fourth equality is direct consequence of 
\begin{align*}
\frac{1}{N}\sum_{j=1}^N \mathbb{E}\{\theta_i\theta_j^\top\} &=\frac{1}{N} \V_{\theta\theta} +\frac{N-1}{N} \U_{\theta\theta} \\ & =\frac{1}{N^2} \sum_{j=1}^N\sum_{t=1}^N \mathbb{E}\{\theta_t\theta_j^\top\}.
\end{align*}
Substituting~\eqref{eqn:exansion:2} into~\eqref{eqn:exansion:1} results in 
\begin{equation*}
\begin{split}
\mathbb{E}\{\|(x+&\theta_i)-\mathbb{E}\{x|y\}\|_2^2\}\\&=\trace(\mathbb{E}\{(x+\theta_i)(x+\theta_i)^\top-(x+\bar{\theta})\mathbb{E}\{x|y\}^\top\\&\hspace{.1in}-\mathbb{E}\{x|y\}(x+\bar{\theta})^\top+ \mathbb{E}\{x|y\}\mathbb{E}\{x|y\}^\top\})\\&
=\mathbb{E}\left\{\|(x+\bar{\theta})-\mathbb{E}\{x|y\}\|_2^2\right\}\\&\hspace{.1in}+\trace\big(\mathbb{E}\{(x+\theta_i)(x+\theta_i)^\top\}\\&\hspace{.2in}-\mathbb{E}\{(x+\bar{\theta})(x+\bar{\theta})^\top\}\big)\\&=\mathbb{E}\left\{\|(x+\bar{\theta})-\mathbb{E}\{x|y\}\|_2^2\right\}\\&\hspace{.1in}+\trace\left(\V_{\theta\theta}-\U_{\theta\theta}\right)(N-1)/N,
\end{split}
\end{equation*}
where the last equality holds due to the fact that
\begin{align*}
\mathbb{E}\{\theta_i\theta_i^\top\}-\mathbb{E}\{\bar{\theta}\bar{\theta}^\top\}&=\V_{\theta\theta}-\frac{1}{N^2}\sum_{j=1}^N\sum_{t=1}^N \mathbb{E}\{\theta_j\theta_t^\top\}\\
&=\V_{\theta\theta}-\frac{1}{N^2}(N\V_{\theta\theta}+(N^2-N)\U_{\theta\theta})\\
&=\frac{N-1}{N}(\V_{\theta\theta}-\U_{\theta\theta}).
\end{align*}
This concludes the proof.

\section{Proof of Corollary~\ref{cor:multiple:herding}}
\label{proof:cor:multiple:herding}
Using Item~(6) in Section 4.2.2~\cite[p.\,50]{lkepohl1996handbook}, we have
\begin{align*}
&\det\left(\matrix{cc}{\lambda I & \frac{1}{\sqrt{N\eta_x\eta_\theta}}I \\ \frac{1}{\sqrt{N\eta_x\eta_\theta}} & (\lambda+\eta_x) I}\right)\\
&\hspace{.4in}=\det(\lambda I) \det\left(\left(\lambda+\eta_x-\frac{1}{N\eta_x\eta_\theta\lambda}\right)I\right)\\
&\hspace{.4in}=\left(\lambda^2+\eta_x\lambda-\frac{1}{N\eta_x\eta_\theta}\right)^{n_x},
\end{align*}
and, as a result, the smallest eigenvalue of the matrix is equal to
$$
\lambda=\frac{-\eta_x-\sqrt{\eta_x^2+4/(N\eta_x\eta_\theta)} }{2}.
$$
For this eigenvalue, we have
\begin{align*}
\matrix{cc}{\lambda I & \frac{1}{\sqrt{N\eta_x\eta_\theta}}I \\ \frac{1}{\sqrt{N\eta_x\eta_\theta}} & (\lambda+\eta_x) I}\matrix{c}{\zeta_1 \\ \zeta_2}=0,
\end{align*}
and, therefore,
$$
\zeta_1=-\frac{1}{\lambda\sqrt{N\eta_x\eta_\theta}}\zeta_2.
$$
On the other hand, we know that $\zeta_1^\top \zeta_1+\zeta_2^\top \zeta_2=1$, which results in
\begin{align*}
\zeta_2
&=\frac{1}{\sqrt{1+1/(\lambda^2\eta_x\eta_\theta N)}}\zeta'\\
&=\sqrt{\frac{N\eta_\theta \eta_x^3+N\eta_\theta \eta_x^2\sqrt{\eta_x^2+4/(N\eta_\theta \eta_x)} +4}{2N\eta_\theta \eta_x^3 + 8}}\zeta',
\end{align*}
for any $\zeta'\in\mathbb{R}^{n_x}$ such that ${\zeta'}^\top \zeta'=1$. Hence,
\begin{align*}
\mathbb{E}\{\|x-&\farhad{[\hat{x}^*(\gamma^*)]}((\gamma^*(x,\theta_i))_{i\in\N})\|_2^2\}\\
&=\mathbb{E}\{\|x-\mathbb{E}\{x|\bar{y}\}\|_2^2\}\\
&=\trace(\V_{xx}-\V_{x\bar{y}}\V_{\bar{y}\bar{y}}^{-1}\V_{\bar{y}x})\\
&=\trace(\V_{xx}-\V_{xx} a^* {a^*}^\top \V_{xx})\\
&=\trace(\V_{xx})-\varsigma(N)\trace(\V_{xx}^{1/2}\zeta' {\zeta'}^\top \V_{xx}^{1/2}),
\end{align*}
where
$$
\varsigma(N)=\frac{N\eta_\theta \eta_x^3+N\eta_\theta \eta_x^2\sqrt{\eta_x^2+4/(N\eta_\theta \eta_x)} +4}{2N\eta_\theta \eta_x^3 + 8}.
$$
Evidently, $\varsigma(N)$ is an increasing function of $N$ because
\begin{align*}
\frac{\mathrm{d}\varsigma(N)}{\mathrm{d}N}
&=\frac{1}{N^2\eta_\theta(\eta_x^2+4/(N\eta_x\eta_\theta))^{3/2})}\\
&\geq 0.
\end{align*}
Thus, $\mathbb{E}\{\|x-\farhad{[\hat{x}^*(\gamma^*)]}((\gamma^*(x,\theta_i))_{i\in\N})\|_2^2\}$ becomes a decreasing function of $N$. In addition, we have
\begin{align*}
\lim_{N\rightarrow \infty}\mathbb{E}\{\|x-\farhad{[\hat{x}^*(\gamma^*)]}((\gamma^*&(x,\theta_i))_{i\in\N})\|_2^2\}\\
&=\farhad{(}\eta_x-\lim_{N\rightarrow \infty}\varsigma(N)\eta_x\farhad{)n_x}=0.
\end{align*}
This concludes the proof.

\end{document}